\documentclass[%
 reprint,
 amsmath,amssymb,
 aps,
]{revtex4-2}

\usepackage{graphicx}
\usepackage{dcolumn}
\usepackage{bm}
\usepackage[utf8]{inputenc}
\usepackage[T1]{fontenc}
\usepackage{mathptmx}
\usepackage{etoolbox}
\usepackage{hyperref}
\usepackage{float}
\usepackage{booktabs}
\usepackage{physics}
\usepackage{color}

\begin{document}

\preprint{APS/123-QED}

\title{Optical Regulation of Chiral-Induced Spin Selectivity}

\author{Wei Liu}
\affiliation{Department of Chemistry, School of Science, Westlake University, Hangzhou 310024 Zhejiang, China}
\affiliation{Institute of Natural Sciences, Westlake Institute for Advanced Study, Hangzhou 310024 Zhejiang, China}
\author{Jingqi Chen}%
\affiliation{Department of Chemistry, School of Science, Westlake University, Hangzhou 310024 Zhejiang, China}
\affiliation{Institute of Natural Sciences, Westlake Institute for Advanced Study, Hangzhou 310024 Zhejiang, China}

\author{Wenjie Dou}
 \email{douwenjie@westlake.edu.cn}

\affiliation{Department of Chemistry, School of Science, Westlake University, Hangzhou 310024 Zhejiang, China}
\affiliation{Department of Physics, School of Science, Westlake University, Hangzhou 310024 Zhejiang, China}
\affiliation{Institute of Natural Sciences, Westlake Institute for Advanced Study, Hangzhou 310024 Zhejiang, China}

\date{\today}

\begin{abstract}
We present a non-perturbative theory that describes how light regulates chiral-induced spin selectivity (CISS) from the perspective of strong light-matter interactions. The research results indicate that 1) light can have opposite effects on the CISS, 2) the difference in CISS is caused by the steady states of nuclei coupled to spin electrons and 3) this steady state differences are caused by the different light-induced Lorentz forces felt by spin-up and spin-down electrons. The fundamental reason for these results is the impact of light on spin-orbital coupling (SOC), which is a complex process. This theoretical framework is verified by the calculations of Floquet SOC non-adiabatic nuclear dynamics.

\end{abstract}

\maketitle
\section{Introduction}
According to Moore's Law\cite{schaller1997moore}, the number of transistors doubles approximately every $18$ to $24$ months, leading to continuous improvement in processor performance and a decrease in chip size. However, as time goes by, Moore's Law faces increasing challenges. Because transistors are already approaching atomic scale, it becomes increasingly difficult to continue reducing their size at the same speed.\cite{lundstrom2003moore,theis2017end} Spintronics has the potential to break through limitations and enable electronic devices to truly reach atomic scale.\cite{shalf2015computing} This multidisciplinary area of study has paved the way for numerous advancements, including the exploration of phenomena such as the spin Hall effect, the utilization of ferroelectric materials, the investigation of topological insulators, and the development of quantum computing technologies.\cite{hirohata2020review}

In recent years, the integration of chiral molecules into the realm of spintronics has opened up new avenues of research and innovation.\cite{naaman2019chiral} Chiral molecules possess asymmetry in their spatial arrangement, resulting in distinct left-handed and right-handed forms.\cite{quack1989structure} This intrinsic chirality can interact with the spin degrees of freedom of electrons, giving rise to a phenomenon known as chiral-induced spin selectivity (CISS), which was initially reported in 1999 by Naaman and collaborators.\cite{ray1999asymmetric}. 
Since then, a number of additional experimental\cite{gohler2011spin,xie2011spin,ben2014local,kiran2016helicenes,eckshtain2016cold,mondal2016photospintronics,alpern2016unconventional,alam2017spin,abendroth2017analyzing,kumar2017chirality,aragones2017measuring,varade2018bacteriorhodopsin,santos2018chirality,gazzotti2018spin,shapira2018unconventional,ghosh2019controlling} and theoretical\cite{yeganeh2009chiral,guo2012spin,gutierrez2012spin,guo2014spin,wu2015spin,medina2015continuum,michaeli2019origin,matityahu2016spin,varela2016effective,pan2016spin,matityahu2017spin,diaz2018effective,diaz2018thermal,maslyuk2018enhanced,nurenberg2019evaluation,yang2019spin} studies have been conducted. The unusual spin polarization observed in CISS has been confirmed to exhibit the following three characteristics: 1) chiral molecules induce spin polarization rather than spin filtering, 2) the direction of spin polarization is determined by the chirality of the molecule and the direction of electron incidence, and 3) the magnitude of spin polarization is dictated by the strength of local spin-orbit coupling (SOC).\cite{wolf2022unusual}

At the microscopic level, we do know of a source of CISS is SOC, but the SOC generated by electron motion within chiral molecules is very weak. Additionally, some spin selection occurs prior to electron passage through the chiral molecule, and this initial spin selection is primarily due to the strong SOC of the metal substrate.\cite{gersten2013induced} How to enhance CISS has attracted extensive research.
Some calculations indicate that electron-vibration coupling enhances the CISS effect, leading to an increase in the electron polarizability in double-helix DNA.\cite{du2020vibration} When considering the coupling between atomic nuclei and electron spins, calculations reveal strong spin selectivity near the conical intersection, even when the SOC is small, with spin selectivity reaching up to $100\%$.\cite{wu2021electronic}
In the presence of a non-zero current, Berry curvature can induce nuclear spin separation and electron spin polarization.\cite{teh2022spin} Even the CISS effect observed in different systems has different mechanisms for amplification.

In this work, we present a non-perturbative theory to describe the regulation of CISS by light from the perspective of strong light-matter interaction. We show three main results, which are important for understanding how to use light to regulate CISS: 1) light can have the opposite effect on the CISS effect, 2) the spin polarization is caused by the different steady states of the motion of nuclei coupled to spin-up and spin-down electrons, and 3) this steady-state difference is caused by the different photoinduced Lorenz-like force felt by electrons with different spins. The underlying cause of these results is the influence of light on SOC, which is a complex process. This theoretical framework is verified by the calculations of Floquet SOC non-adiabatic nuclear dynamics.

\section{Theory}
We start from a simple model with two spatial orbitals coupled to left and right leads whose voltages are $\mu_L$ and $\mu_R$, such that the Hamiltonian depends on two nuclear degrees of freedom (x and y are considered, uniformly represented as $\mathbf R$). The total model Hamiltonian $\hat H_{tot}$ consists of the kinetic energy and another three parts: the system Hamiltonian $\hat H_s$, the bath Hamiltonian $\hat H_b$ composed of left and right leads, and the system-bath coupling Hamiltonian $\hat H_{sb}$,  
\begin{eqnarray}
&\hat{H}_{tot} = \frac{P^2}{2M}+\hat H_s  + \hat H_b + \hat H_{sb},   \\
&\hat H_s  = \sum_{ij} [h_s]_{ij} (\mathbf R, t) \hat d_i^\dagger \hat d_j + U(\mathbf R), \\
&\hat H_b = \sum_{k\zeta } \epsilon_{k\zeta  } \hat c_{k\zeta}^\dagger \hat c_{k\zeta },  \\
&\hat H_{sb}  =  \sum_{\zeta k,i} V_{\zeta k,i}  (   \hat c_{k\zeta}^\dagger \hat d_i + \hat d^\dagger_i  \hat c_{k\zeta}  ).\label{eqn:sb}
\end{eqnarray}
In system Hamiltonian ($H_s$), the operators $\hat d_i^\dagger$ and $\hat d_i$ respectively create and annihilate an electron in the $i$-th spin orbital of the subsystem, while $U(\mathbf R)$ represents the electrostatic potential between nuclei. As for the Hamiltonian ($H_b$), the operators $\hat c_{k\zeta}^\dagger$ and $\hat c_{k\zeta}$ create and annihilate electrons in the $k$-th spin orbital of the lead $\zeta$, where the energy of the orbital is denoted by $\epsilon_{k\zeta}$. Lastly, in the interaction Hamiltonian ($H_{sb}$), the tunneling element $V_{\zeta k,i}$ indicates the interaction between subsystem spin orbital $i$ and lead spin orbital $k\zeta$. In Eq.~\ref{eqn:sb}, the Condon approximation has been utilized, implying that $V_{\zeta k,i}$ is independent of $\mathbf R$.

Next, we focus on a shifted parabola model under periodic driving:
\begin{equation}
h_s(\mathbf R,t) = [Ax+C\cos(\Omega t)]\sigma_x + By\sigma_y + (x+\Delta)\sigma_z.
\label{eqn:hs}
\end{equation}
Here, $\sigma_x$, $\sigma_y$, and $\sigma_z$ represent Pauli matrices, respectively. $C$ and $\Omega$ denote the amplitude and frequency of the periodic driving, respectively. $\Delta$ represents the energy gap between two spatial orbitals, which is zero as to symmetric spatial orbitals and non-zero as to asymmetric spatial orbitals. Moreover, parameters A, B are introduced to control the rates of diabatic, spin-orbit with respect to geometries x and y. The shifted parabola model is extensively employed to simulate electron transfer and excitation energy transfer processes.\cite{nitzan2006chemical} Note that Eq.~\ref{eqn:hs} is an effective Hamiltonian, where positive $B$ represents spin-up and negative $B$ represents spin-down.

Furthermore, we present the scalar potential $U(\mathbf R)$ that characterizes the tilted energy landscape, illustrated below:\cite{malhado2012photoisomerization}
\begin{equation}
U(\mathbf R) = \frac{1}{2}(x - \lambda_x{x})^2 + \frac{1}{2}(y - \lambda_y{y})^2.
\end{equation}
Note that the scalar potential $U(\mathbf R)$ is not affected by electronic friction. Additionally, the linear terms $\lambda_x$ and $\lambda_y$ serve to introduce asymmetry to the adiabatic state, and we can adjust them to generate the strong spin current necessary for our purposes. 
At this point, our model has been fully introduced. We use Floquet theory and non equilibrium Green's function method to solve this model, and the specific solution process is listed in the Appendix~\ref{sec:EFmodel}.

Finally, we define spin polarization $\xi$ as follows, 
\begin{eqnarray}
\xi = \frac{I^{\downarrow} - I^{\uparrow}}{I^{\downarrow} + I^{\uparrow}} \times 100 \%.
\label{eqn:sp}
\end{eqnarray}
Here, $I^{\uparrow/\downarrow}$ denotes spin-up or spin-down current. The specifc procedures of solving $I^{\uparrow/\downarrow}$ see Appendix~\ref{sec:cspin}.

\section{Results and Discussions}
\label{sec:results}
\subsection{Spin Polarization Manipulated by Light}
\begin{figure}[htbp]
\includegraphics[width=.48\textwidth]{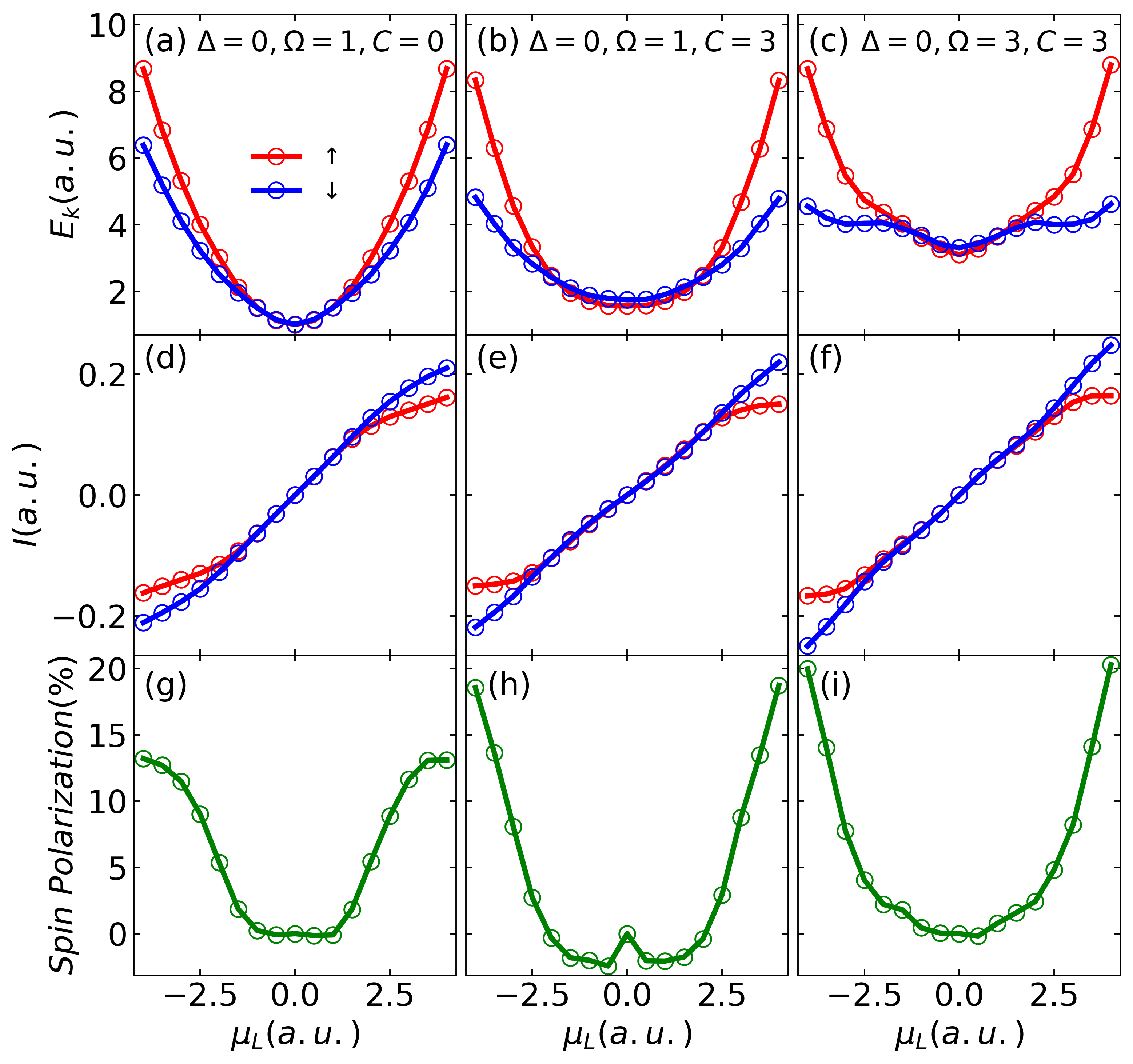}
\caption{We consider a system with parameters $\Delta=0$, $A=1$, $B=1.2$, $kT = 0.5$, $\lambda_x =0$, $\lambda_y =1$, $\widetilde{\Gamma} = 1$, $\mu_L = -\mu_R$ and Floquet level $N=5$, where $C$ takes the value of 0 for cases (a), (d), and (g), 3 with $\Omega = 1$ for cases (b), (e), and (h), and 3 with $\Omega = 3$ for cases (c), (f), and (i). Using Eq.~\ref{eqn:iupdown} and Eq.~\ref{eqn:sp}, we calculate the spin current and spin polarization. No spin polarization is expected at zero voltage bias, while a significant polarization can be observed for finite values of $\mu_L$.}
\label{fig1}
\end{figure}

\begin{figure}[htbp]
\includegraphics[width=.48\textwidth]{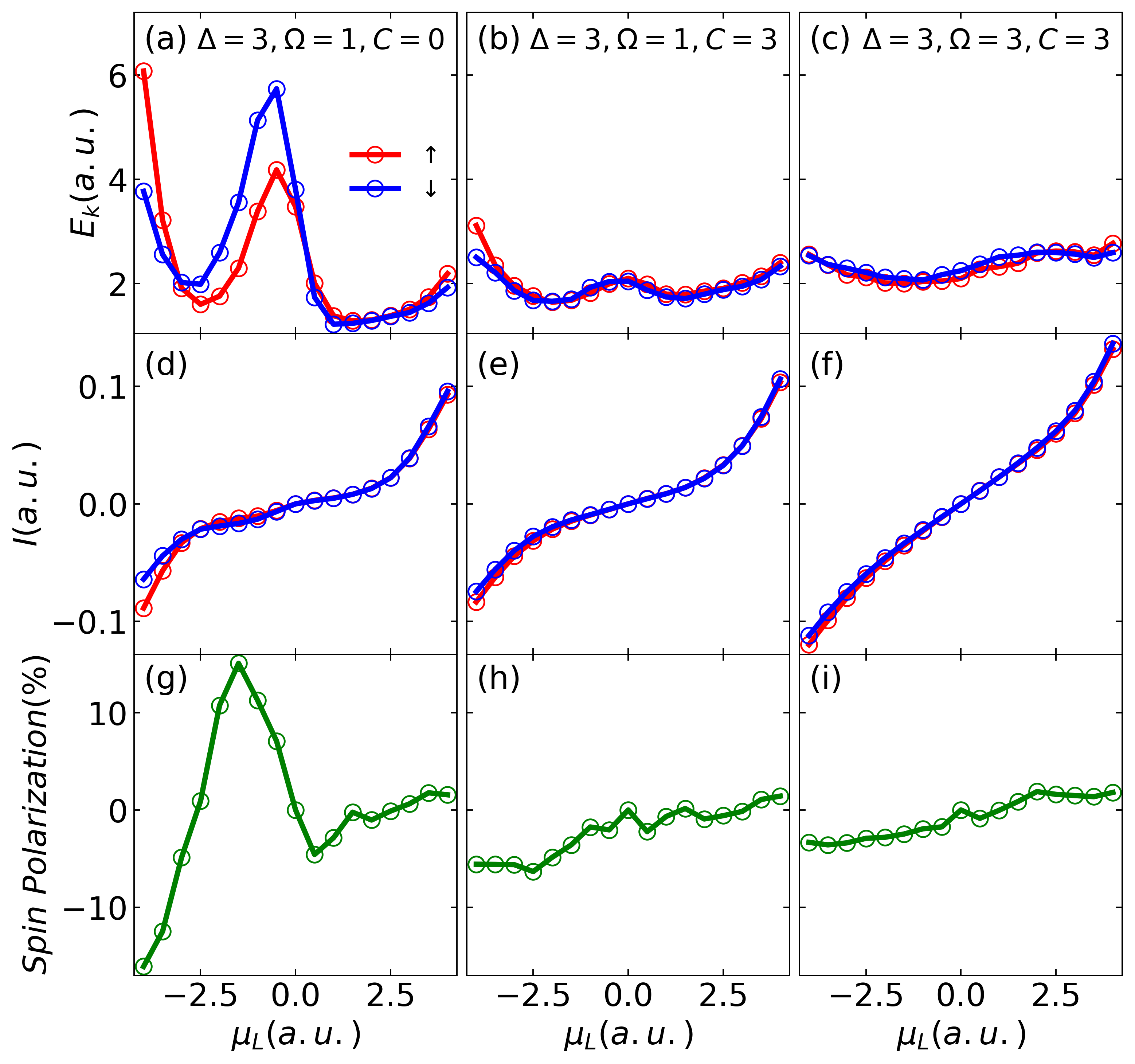}
\caption{We consider a system with parameters $\Delta=3$, $A=1$, $B=1.2$, $kT = 0.5$, $\lambda_x =0$, $\lambda_y =0.8$, $\widetilde{\Gamma} = 1$, $\mu_L = -\mu_R$ and Floquet level $N=5$, where $C$ takes the value of 0 for cases (a), (d), and (g), 3 with $\Omega = 1$ for cases (b), (e), and (h), and 3 with $\Omega = 3$ for cases (c), (f), and (i).}
\label{fig2}
\end{figure}

In this section, we present our findings on how light manipulation affects spin polarization. We provide steady-state kinetic energy, spin current, and spin polarization diagrams in Fig.~\ref{fig1} for symmetric ($\Delta=0$) and in Fig.~\ref{fig2} for asymmetric ($\Delta=3$) cases both with $\mu_L = -\mu_R$.

In Fig.~\ref{fig1}, the left panel [(a), (d), and (g)] represents cases without light ($C=0$), where the CISS effect induces spin polarization that increases with the absolute value of $\mu_L$. Specifically, for the symmetric case, the increase in spin polarization as $\mu_L$ becomes more negative tends to converge. When we introduce light regulation in the middle panel [(b), (e), and (h)] with $\Omega=1$ and $C=3$, we observe a significant increase in spin polarization, particularly when $\mu_L=4.0$, where spin polarization increases from $13.1\%$ to $18.7\%$ compared to the non-light case. Further increasing $\Omega$ to $3$ in the right panel [(c), (f), and (i)] results in a continued increase in spin polarization, reaching $20.3\%$. These results indicate that, in the symmetric case, light significantly enhances spin polarization, regardless of increasing light intensity or frequency.

In Fig.~\ref{fig2}, we investigate the asymmetric case ($\Delta=3$) with $\mu_L = -\mu_R$, revealing an opposite phenomenon. The left panel [(a), (d), and (g)] represents cases without light ($C=0$), where spin polarization is observed due to the CISS effect. However, in the middle panel [(b), (e), and (h)], the introduction of light regulation with $\Omega=1$ and $C=3$ results in a decrease in spin polarization, particularly when $\mu_L=-4.0$, where spin polarization decreases from $-16.1\%$ to $-5.58\%$ compared to the non-light case. Further increasing $\Omega$ to $3$ in the right panel [(c), (f), and (i)] leads to a further suppression of spin polarization to $-3.33\%$. These findings demonstrate that, in asymmetric scenarios, the presence of light significantly suppresses spin polarization, with the suppression effect becoming more pronounced as light intensity increases and frequency rises.

It is worth noting that the spatial distribution of current remains unaffected by spin, while spin polarization undergoes a completely opposite trend. To clarify this observation, we turn our attention to the distinct steady states of nuclear motion associated with electrons of different spin directions in both symmetric and asymmetric cases. This crucial distinction is further elaborated upon in the subsequent sections.

\subsection{Steady States Manipulated by Lorentz-like Force}
\label{sec:spss}

\begin{figure}[htbp]
\includegraphics[width=.48\textwidth]{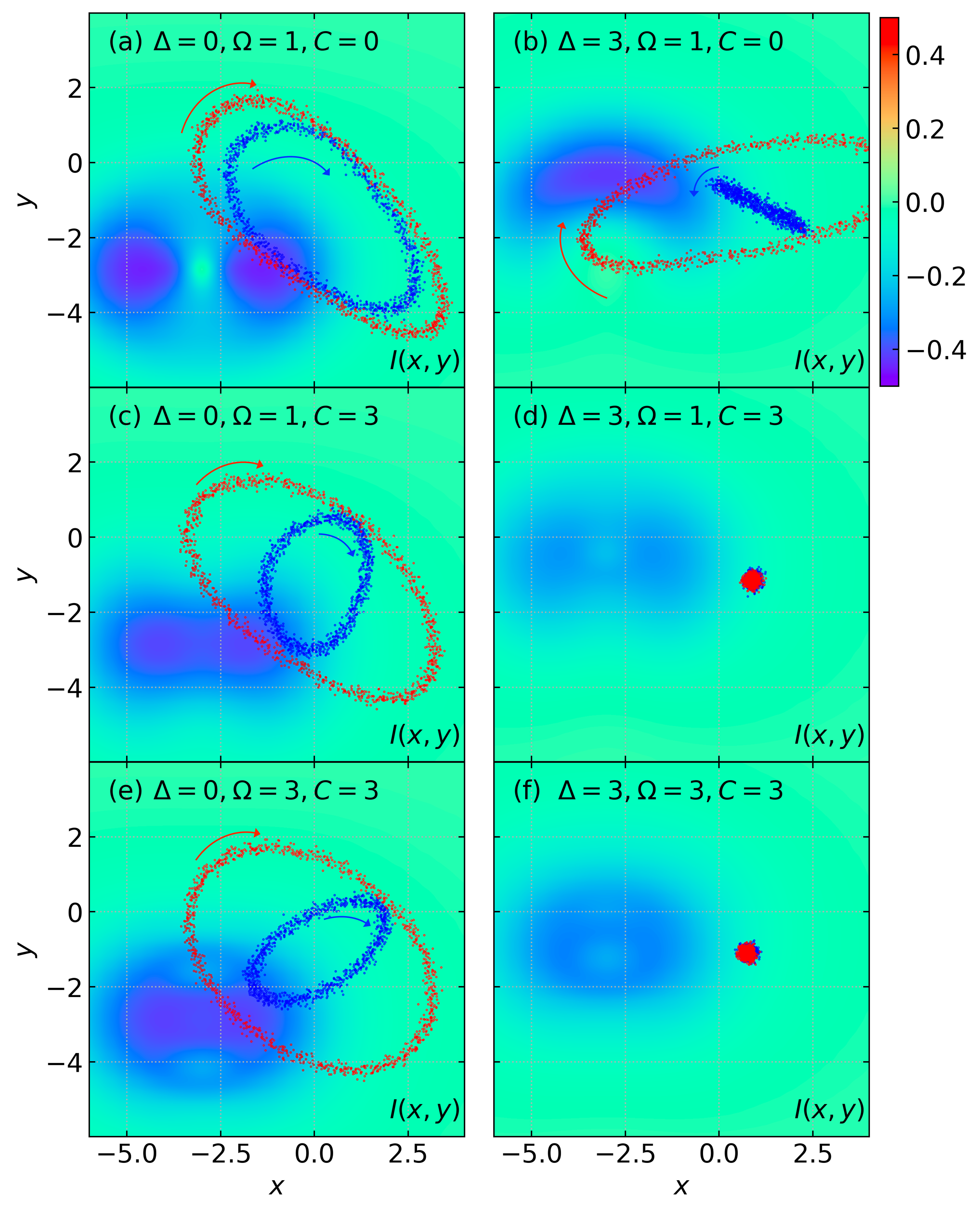}
\caption{The current distribution is depicted by a heat map, where the red dot indicates the steady states of nuclei with spin up electrons and the blue dot represents the steady states of nuclei with spin down electrons. The values of the general parameters used in this figure include $\mu_R=-\mu_L$, $A=1$, $B=1.2$, $kT=0.5$, $\lambda_x=0$, $\widetilde{\Gamma}=1$ and Floquet level $N=5$. For cases (a), (c), and (e), the parameters employed are as follows: $\Delta=0$, $\mu_L=4.0$ and $\lambda_y=1$. On the other hand, for cases (b), (d), and (f), the parameters used include $\Delta=3$, $\mu_L=-4.0$ and $\lambda_y=0.8$. It is worth noting that for cases (a) and (b), the value of the parameter $C$ is set to 0. In contrast, for cases (c) and (d), the value of $C$ takes on a value of 3 with $\Omega=1$. Lastly, for cases (e) and (f), the parameter $C$ is also set to 3 but with $\Omega=3$.}
\label{fig3}
\end{figure}

The observed different steady states of nuclei with spin-up and spin-down electrons, as depicted in Fig.~\ref{fig3}, can be attributed to the influence of the Lorentz-like force derived from $\gamma_{xy}^{A}$.\cite{mosallanejad2023floquet} This force primarily acts in the $\gamma_{xy}^{A}<0$ region, inducing rotational motion in a clockwise direction for both spin directions. However, the rotations in nuclei's steady states are absent in the asymmetric case of light regulation, mainly due to electronic friction exceeding excitation, causing all nuclei to dissipate to the same point.

\begin{figure}[htbp]
\includegraphics[width=.48\textwidth]{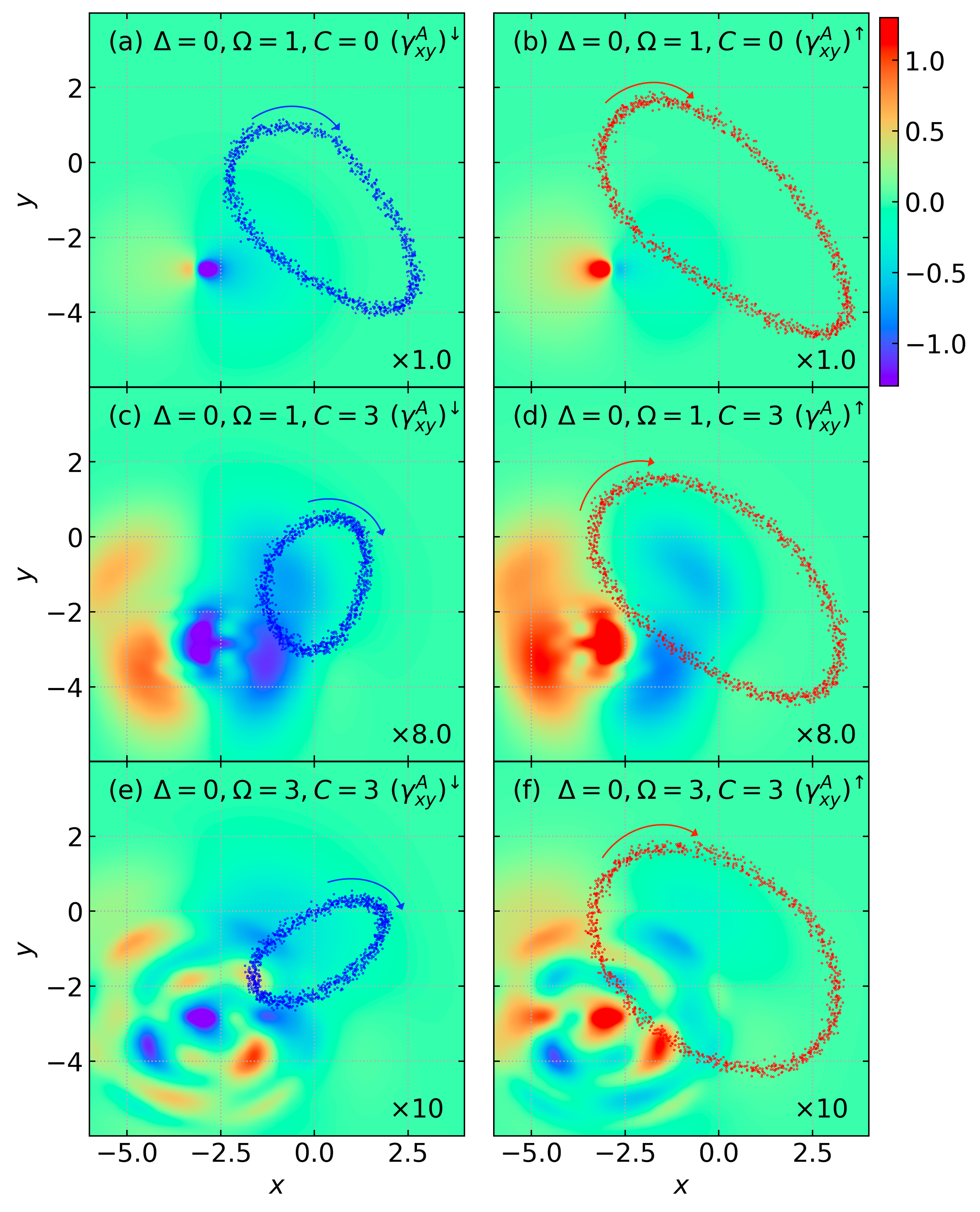}
\caption{$\gamma_{xy}^{A}$ is depicted by a heat map, where the red and blue dots represent the steady states of nuclei with spin-up and spin-down electrons. The values of the parameters used in this figure include $\Delta=0$, $\mu_L=4.0$, $\mu_R=-\mu_L$, $A=1$, $B=1.2$, $kT=0.5$, $\lambda_x=0$, $\lambda_y=1$, $\widetilde{\Gamma}=1$ and Floquet level $N=5$. It is worth noting that for cases (a) and (b), the value of the parameter $C$ is set to 0. In contrast, for cases (c) and (d), the value of $C$ takes on a value of 3 with $\Omega=1$. Lastly, for cases (e) and (f), the parameter $C$ is also set to 3 but with $\Omega=3$.}
\label{fig4}
\end{figure}

In Fig.~\ref{fig4}, we visualize $\gamma_{xy}^{A}$ for the symmetric case ($\Delta=0$). The red and blue dots represent the steady states of nuclei with spin-up and spin-down electrons, respectively. Both spin directions experience clockwise rotations due to the negative values of $\gamma_{xy}^{A}$, explaining the observed spin polarization in this scenario.

\begin{figure}[htbp]
\includegraphics[width=.48\textwidth]{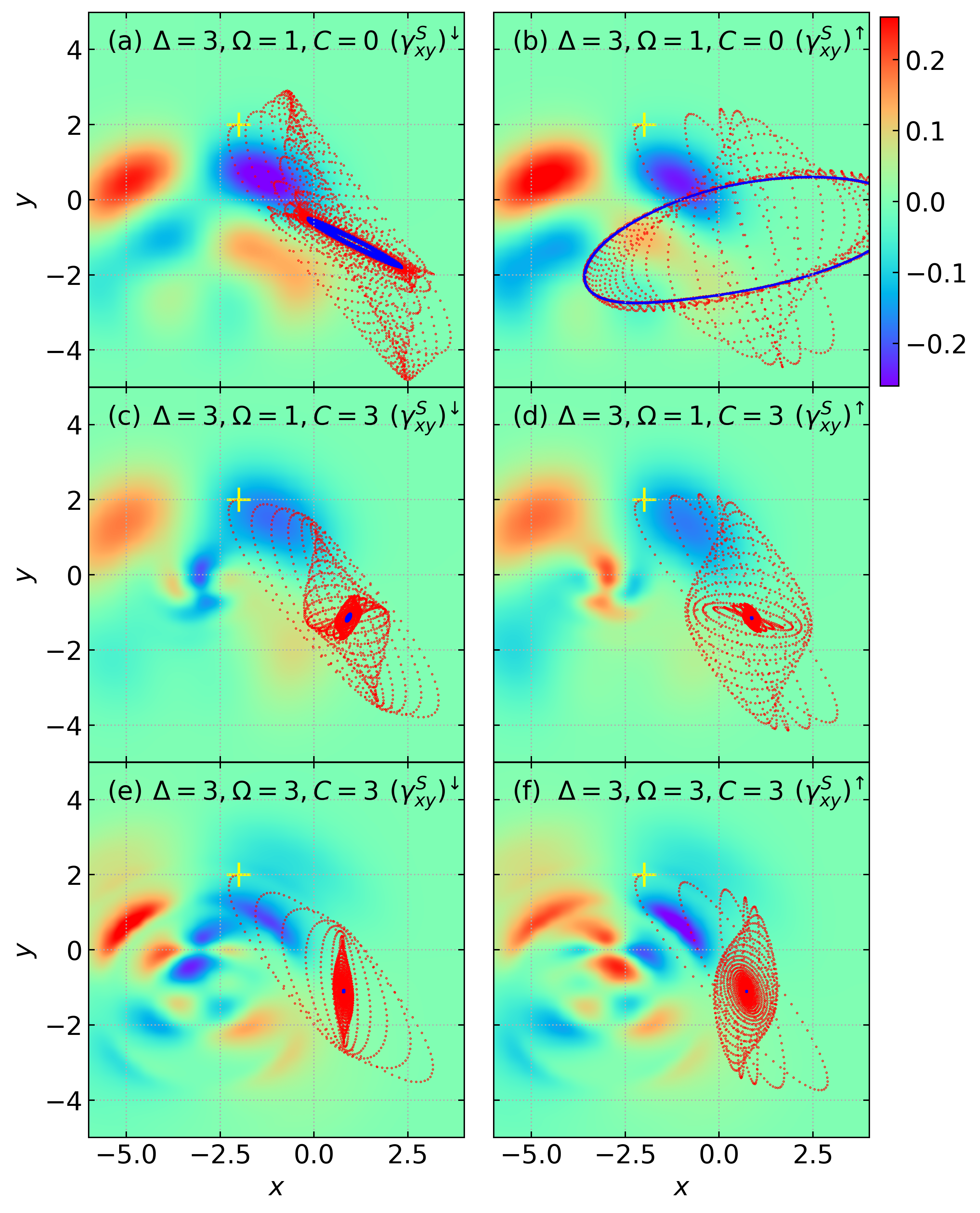}
\caption{The heat map illustrates $\gamma_{xy}^{S}$, where the red and blue dots indicate the unsteady and steady states of nuclei, respectively. The yellow cross marks the initial position of all six cases, which are $(-2, 2)$ and initial momentum $(p_x=-1, p_y=0)$. The parameters used in this figure are $\Delta=3$, $\mu_L=-4.0$, $\mu_R=-\mu_L$, $A=1$, $B=1.2$, $kT=0.5$, $\lambda_x=0$, $\lambda_y=0.8$, $\widetilde{\Gamma}=1$, and Floquet level $N=5$. It's important to note that for cases (a) and (b), the parameter $C$ is set at 0, while for cases (c) and (d), $C$ takes a value of 3 with $\Omega=1$. Lastly, for cases (e) and (f), the parameter $C$ is again set to 3, but with $\Omega=3$.}
\label{fig5}
\end{figure}

However, in the asymmetric case with light regulation, the predominant effect of $\gamma_{xy}^{S}$,\cite{mosallanejad2023floquet} as shown in Fig.~\ref{fig5}, leads to either energy injection or dissipation from the environment, preventing the observed rotations and spin polarization.

In Fig.~\ref{fig5}, we present $\gamma_{xy}^{S}$ for the asymmetric case ($\Delta=3$), where the red and blue dots represent the unsteady and steady states of nuclei, respectively. The predominance of $\gamma_{xy}^{S}>0$ results in continuous energy dissipation from the environment, ultimately preventing rotations and causing a lack of spin polarization, as observed in this scenario.

These findings highlight the critical role of the Lorentz-like force in influencing the steady states of nuclear motion and, consequently, spin polarization in electronic systems subjected to light regulation.

\subsection{Broader Implications}
In addition to the specific cases discussed here, our results have broader implications for the fields of spintronics and optoelectronics. The ability to manipulate spin polarization through light regulation opens up possibilities for developing spin-based devices with tunable properties. Furthermore, the understanding of the underlying mechanisms, such as the influence of Lorentz-like forces, can guide the design of novel materials and structures for spin control and manipulation.

\section{Conclusions}
\label{sec:conclu}
In conclusion, our investigation into the influence of light regulation on spin polarization in electronic systems has yielded valuable insights. We have demonstrated that light can exert a profound impact on spin polarization, with its effects varying depending on the system's symmetry and the characteristics of the light source. Our study has also shed light on the underlying mechanisms at play, particularly the pivotal role of Lorentz-like forces in manipulating the steady states of nuclear motion and thereby influencing spin polarization.

These findings not only advance our fundamental understanding of spin dynamics in optoelectronic systems but also open up exciting opportunities for practical applications in the realm of spintronics. The ability to control and manipulate spin polarization through light-matter interactions holds great promise for the development of novel technologies, including spin-based quantum computing and information storage.

As we peer into the future, several promising avenues for further research emerge. In the realm of electronic structure theory, there is an urgent need for higher-precision calculations of excited-state electronic spin states in large systems, surpassing the capabilities of traditional Density Functional Theory (DFT). Accurately modeling spin-orbit coupling (SOC) and its derivative couplings is equally crucial, as SOC plays a pivotal role in shaping electronic spin states and interactions.

Moving forward, it is imperative to validate our findings by applying first-principles calculations to real-world systems, transcending the confines of our model. The evidence presented here suggests that we are on the cusp of an era where spin polarization manipulation through light-matter interactions becomes increasingly feasible. We are optimistic that our work will have far-reaching implications for a multitude of disciplines, propelling the fields of spintronics, quantum computing, and beyond, and ultimately enabling the breakthroughs needed to surpass current limits in information storage and computing power.

This work is supported by the startup funding from Westlake University. 
W.L. acknowledges support from the high-performance computing center of Westlake University.


\appendix
\section{Floquet Electronic Friction Model}
\label{sec:EFmodel}
The periodic driving term (Eq.~\ref{eqn:hs}) makes the system Hamiltonian $H_s$ time dependent, posing difficulties in numerically solving the model. Electronic friction could be evaluated by the Green's function methods when $H_s$ is time independent, however, previous Green's function cannot deal with time dependent Hamiltonian \cite{dou2017born}. To tackle the complex and recurring time-dependent issue at hand, we have recently devised a Floquet electronic friction model that employs non-equilibrium Green's functions \cite{mosallanejad2023floquet}.

In Floquet electronic friction model, the dynamics of the molecule in the junction could be described by the following Langevin dynamics:
\begin{equation}
  M_\mu \ddot R_\mu = F_\mu^F - \sum_\nu \gamma_{\mu\nu}^F \dot R_\nu +  \delta F_\mu^F,
\label{eqn:langevin}
\end{equation}
where $M$ and $R$ refer to the mass and position of the nuclei, while $\mu$ and $\nu$ symbolize the degrees of freedom (DOFss); the superscript $F$ denotes the term after the Floquet transformation. Concerning the right-hand side of the equation, $F_\mu^F$ represents the mean force, $\gamma_{\mu\nu}^F$ stands for the friction tensor, and $\delta F_{\mu}^F$ denotes a Markovian random force \cite{bode2012current, smith1993electronic, lu2012current, chen2018current}. The equation at hand can be interpreted as follows: the motion of nuclei disrupts the equilibrium of the system's electronic structure. Given a prompt electronic response, the electrons generate an opposing force that responds to the nuclear movements.

The feedback between the nuclei and electrons can be simulated through approximations of frictional damping and Markovian random force \cite{dou2015frictional,dou2017born}. The following expressions are used to denote the adiabatic force ($F^F_{\mu}$), friction tensor ($\gamma_{\mu\nu}^F$), and the symmetrized correlation function of the random force [$\frac{1}{2}(D_{\mu\nu}^F + D_{\nu\mu}^F)$]:
\begin{equation}
F^F_{\mu}=  -\frac{1}{2\pi i(2N+1)} \\
 \int_{-\infty}^{+\infty}d\epsilon Tr\left\{ \partial_{\mu} h_s^F  G_{<}^F \right\} - \partial_{\mu} U,\\
\end{equation}
\begin{equation}
\gamma_{\mu\nu}^F = - \frac{\hbar}{2\pi (2N+1)}\\
\int_{-\infty}^{+\infty}d\epsilon Tr\left \{ \partial_{\mu} h_s^F \partial_{\epsilon} G_r^F \partial_{\nu} h_s^F G_{<}^F \right \} + H.c.,\\
\end{equation}
\begin{equation}
\frac{1}{2}(D_{\mu\nu}^F + D_{\nu\mu}^F) =  \frac{\hbar}{4\pi (2N+1)} \\
\int_{-\infty}^{+\infty}d\epsilon Tr\left\{ \partial_\mu h_s^F  G_{>}^F \partial_\nu h_s^F  G_{<}^F \right\},
\label{eqn:random}
\end{equation}
where $G_r^F$, $G_a^{F}$, $G_{<}^F$, and $G_{>}^F$ are the retarded, advanced, lesser, and larger Floquet Green’s functions of the electron in the energy domain, respectively. The Floquet Green's function is given by 
\begin{eqnarray}
\begin{aligned}
\label{e19}
G_r^{F}(\epsilon) &= \left(  \epsilon-\Sigma_r^F -h_s^F \right) ^{-1},\\
G^F_{<,>}(\epsilon) &= G^F_{r} (\epsilon) \Sigma^F_{<,>} (\epsilon)   G^F_{a} (\epsilon),\\
G_a^{F}(\epsilon) &= G_r^{F}(\epsilon)^\dagger,
\end{aligned}
\end{eqnarray}
where 
\begin{eqnarray}
\begin{aligned}
\label{e20}
\Sigma_r^F &= - \frac{i}{2} \widetilde{\Gamma},\\
\Sigma^{F}_{<} &= i\widetilde{\Gamma}\sum_{n = -N}^{N} 
\begin{pmatrix}
     f_{n,L} & 0  \\
     0   & f_{n,R}
\end{pmatrix} \hat{L}_n,
\\
\Sigma^{F}_{>} &= -i\widetilde{\Gamma}\sum_{n = -N}^{N}
\begin{pmatrix}
     1-f_{n,L} & 0  \\
     0   & 1-f_{n,R}
\end{pmatrix} \hat{L}_n,\\
\end{aligned}
\end{eqnarray}
and
\begin{eqnarray}
\begin{aligned}
&f_{n,L} = f(\epsilon+n\hbar\omega,\mu_L),\\
&f_{n,R} = f(\epsilon+n\hbar\omega,\mu_R).
\end{aligned}
\end{eqnarray}
Here, according to the wide-band approximation, $\widetilde{\Gamma}$ is a constant representing the coupling strength between the system and the bath, $\hat{L}_n$ is the n-th ladder operator which turns the vector-like Fourier expansion into a matrix-like representation, $\epsilon$ is called quasi-energy and $f$ is Fermi function, 
\begin{equation} 
f(\epsilon, u) = \frac{1}{\exp[\beta(\epsilon-u)]+1},
\end{equation}
where $\beta=1/kT$.

In particular, we define the symmetric and antisymmetric components $[\gamma_{\mu\nu}^{S} = (\gamma_{\mu\nu}^{F} + \gamma_{\nu\mu}^{F})/2, \gamma_{\mu\nu}^{A} = (\gamma_{\mu\nu}^{F} - \gamma_{\nu\mu}^{F})/2]$ of the friction tensor \cite{mosallanejad2023floquet}. The symmetric component $\gamma_{\mu\nu}^{S}$ of the friction tensor has contrasting effects depending on whether it has a positive or negative value. For example, when it is positive, it dissipates energy, whereas a negative value implies it can excite energy \cite{mosallanejad2023floquet, teh2021antisymmetric}. The antisymmetric component $\gamma_{\mu\nu}^{A}$ of the friction tensor for electrons is associated with a non-zero Berry curvature. This phenomenon implies that the nuclear wave packets are subject to different pseudo magnetic fields, which can lead to either a clockwise or counterclockwise influence on the direction of steady-state electron rotation \cite{mosallanejad2023floquet, teh2021antisymmetric}. 

During our study, all terms were expressed using the Floquet transformation. It is essential to note that the total number of Floquet levels is equal to $(2N+1)$, where the precise value of $N$ must be determined to obtain reliable outcomes. In particular, we need to choose $N$ which is large enough to converge the current numerically. (see the procedures for calculating current in Sec.~\ref{sec:cspin}.)

To evaluate the the adiabatic force ($F^F_{\mu}$), friction tensor ($\gamma_{\mu\nu}^F$), and the symmetrized correlation functions [$\frac{1}{2}(D_{\mu\nu}^F + D_{\nu\mu}^F)$], we need to represent $h_s(\mathbf R,t)$ (Eq.~\ref{eqn:hs}) in Floquet levels.
Here, we provide an explicit formula for each block of the extensive matrix, which is in the Fourier space:
\begin{eqnarray}
h_s^F = \sum_{n = -N}^{N} h_s^{(n)}\hat{L}_n + \hat{N}\otimes\hat{I}_n \hbar \omega,
\label{eqn:hsF}
\end{eqnarray}
\begin{eqnarray}
h_s^{(n)} = \frac{1}{T}\int_{0}^{T}h_s(\mathbf R,t)e^{-in\Omega 
 t}dt,
\label{eqn:hsn}
\end{eqnarray}
where $\hat{N}$ is the number operator, $\hat{I}_n$ is the identity matrix for the $n$-th order and $T$ is the period, equal to $2\pi/\Omega$. Specifically, Eq.~\ref{eqn:hsF} represents the decomposition of $h_s(\mathbf R,t)$ into different energy components rotating at different frequencies in the Fourier space. The term $h_s^{(n)}$ denotes the decomposition of $h_s(\mathbf R,t)$ at the frequency of $n\Omega$. The first term of Eq.~\ref{eqn:hsF} is the weighted sum of each Floquet potential, while the second term is the tensor product of $\hat{N}$ and $\hat{I}_n$. This term can be added to the Hamiltonian because $\hat{N}$ commutes with the Hamiltonian. Eq.~\ref{eqn:hsF} allows us to understand the time evolution of the system at different rotating frequencies, and further investigate the physical properties of the system.

Lastly, we will outline how to generate the random force ($\delta F_{\mu}^{F}$) in the Langevin dynamics in Eq.~\ref{eqn:langevin}. It is essential to emphasize that the matrix ${D}_{\mu\nu}^F$ in Eq.~\ref{eqn:random} specifically represents the correlation function matrix of a randomly generated force. However, we have not yet provided a clear explanation of how to generate random forces that correspond to various DOFs.
Within the context of the Hamiltonian being studied, only two DOFs exist, namely $x$ and $y$. As a result, our correlation function matrix can be expressed as:
\begin{eqnarray}
D^F = 
    \begin{pmatrix}
      D_{xx}^F &  D_{xy}^F \\
      D_{yx}^F &  D_{yy}^F
    \end{pmatrix}.
\end{eqnarray}

We shall proceed by diagonalizing this matrix to obtain the diagonal matrix, denoted as:
\begin{eqnarray}
 \widetilde{D}^F = 
    \begin{pmatrix}
      \widetilde{D}_{x}^F & 0 \\
      0  &  \widetilde {D}_{y}^F
    \end{pmatrix}.
\end{eqnarray}

Subsequently, we generate two random numbers following a Gaussian distribution using the norm $\sigma_\mu = \sqrt{2\widetilde{D}_\mu^F/dt}$, where $\mu = x, y$. Lastly, to obtain the random forces that correspond to the $x$ and $y$ directions, we multiply the eigenvectors of $D^F$ by the random number matrix.

\section{Evaluate the Spin Current}
\label{sec:cspin}
In this section, we provide a detailed account of the current and spin polarization calculations. 

First, we outline how to calculate the spin current.In the molecular junction, we consider a molecule connected with a left and a right metal electrode, with chemical potential $\mu_L$ and $\mu_R$ , respectively. The two electrode will contribute to a voltage $V=\mu_R - \mu_L$. The spin current $I^{\uparrow/\downarrow}$ is obtained by averaging over the nuclear DOFs through Langevin dynamics, and is given by the expression \cite{teh2022spin}:
\begin{eqnarray}
I^{\uparrow/\downarrow} = \int d\mathbf R d\mathbf P I_{loc}^{F}(\mathbf R) \rho^{\uparrow/\downarrow}(\mathbf R, \mathbf P). 
\label{eqn:iupdown}
\end{eqnarray}
Here, the probability density distribution of nuclear configuration $\rho^{\uparrow/\downarrow}(\mathbf R, \mathbf P)$ is obtained by sampling 10000 trajectories according to Eq.~\ref{eqn:langevin}, and evaluating the steady-state phase space trajectories. We make the ansatz that the Floquet local spin current $I_{loc}^{F}$ flowing from the left lead through the molecule to the right lead can be evaluated by the Landauer formula \cite{haug2008quantum, meir1992landauer},
\begin{eqnarray}
\begin{aligned}
I_{loc}^F &=\\
&\frac{e}{2\pi\hbar(2N+1)} \int_{-\infty}^{+\infty}d\epsilon Tr \left \{ T^F(\epsilon)[f_L^F(\epsilon)-f_R^F(\epsilon)] \right \}. \\
\end{aligned}
\label{eqn:iloc}
\end{eqnarray}
In the above equation, $f_L^F(\epsilon)$ and $f_R^F(\epsilon)$ represent the Floquet Fermi-Dirac distribution functions that correspond to the left and right leads, respectively. Additionally, $T^F(\epsilon)$ denotes the Floquet transmission probability, which can be deconstructed using the Green's functions.
\begin{eqnarray}
T^F(\epsilon) = \Gamma_L^F G^F_r(\epsilon) \Gamma_R^F G_a^F(\epsilon),
\label{eqn:tran}
\end{eqnarray}
where
\begin{eqnarray}
\begin{aligned}
\Gamma_L^F = \Gamma_L \otimes \hat{I}_n, \quad
\Gamma_R^F = \Gamma_R \otimes \hat{I}_n.
\end{aligned}
\end{eqnarray}

In this project, we examine the distinct interaction between a molecule and two leads where solely orbital 1 ($x+\Delta$) couples to the left lead and only orbital 2 ($-x-\Delta$) couples to the right lead. This arrangement is demonstrated in the subsequent definition of the $\Gamma$ matrices:
\begin{eqnarray}
 \Gamma_L = 
    \begin{pmatrix}
      \widetilde{\Gamma} & 0 \\
      0  &  0
    \end{pmatrix},\quad
 \Gamma_R = 
    \begin{pmatrix}
       0& 0 \\
      0  &  \widetilde{\Gamma}
    \end{pmatrix}.
\end{eqnarray}

Note that $T^F(\epsilon)$ is invariant to changing $B \rightarrow -B$, which implies that the local current $I_{loc}^{F}$ is in fact independent of the exact spin carrier. For this reason we have not included any superscripts $\uparrow/\downarrow$ in Eq.~\ref{eqn:iloc} and Eq.~\ref{eqn:tran}.

\nocite{*}

\bibliography{ref}

\begin{thebibliography}{60}%
\makeatletter
\providecommand \@ifxundefined [1]{%
 \@ifx{#1\undefined}
}%
\providecommand \@ifnum [1]{%
 \ifnum #1\expandafter \@firstoftwo
 \else \expandafter \@secondoftwo
 \fi
}%
\providecommand \@ifx [1]{%
 \ifx #1\expandafter \@firstoftwo
 \else \expandafter \@secondoftwo
 \fi
}%
\providecommand \natexlab [1]{#1}%
\providecommand \enquote  [1]{``#1''}%
\providecommand \bibnamefont  [1]{#1}%
\providecommand \bibfnamefont [1]{#1}%
\providecommand \citenamefont [1]{#1}%
\providecommand \href@noop [0]{\@secondoftwo}%
\providecommand \href [0]{\begingroup \@sanitize@url \@href}%
\providecommand \@href[1]{\@@startlink{#1}\@@href}%
\providecommand \@@href[1]{\endgroup#1\@@endlink}%
\providecommand \@sanitize@url [0]{\catcode `\\12\catcode `\$12\catcode `\&12\catcode `\#12\catcode `\^12\catcode `\_12\catcode `\%12\relax}%
\providecommand \@@startlink[1]{}%
\providecommand \@@endlink[0]{}%
\providecommand \url  [0]{\begingroup\@sanitize@url \@url }%
\providecommand \@url [1]{\endgroup\@href {#1}{\urlprefix }}%
\providecommand \urlprefix  [0]{URL }%
\providecommand \Eprint [0]{\href }%
\providecommand \doibase [0]{https://doi.org/}%
\providecommand \selectlanguage [0]{\@gobble}%
\providecommand \bibinfo  [0]{\@secondoftwo}%
\providecommand \bibfield  [0]{\@secondoftwo}%
\providecommand \translation [1]{[#1]}%
\providecommand \BibitemOpen [0]{}%
\providecommand \bibitemStop [0]{}%
\providecommand \bibitemNoStop [0]{.\EOS\space}%
\providecommand \EOS [0]{\spacefactor3000\relax}%
\providecommand \BibitemShut  [1]{\csname bibitem#1\endcsname}%
\let\auto@bib@innerbib\@empty
\bibitem [{\citenamefont {Schaller}(1997)}]{schaller1997moore}%
  \BibitemOpen
  \bibfield  {author} {\bibinfo {author} {\bibfnamefont {R.~R.}\ \bibnamefont {Schaller}},\ }\bibfield  {title} {\bibinfo {title} {Moore's law: past, present and future},\ }\href@noop {} {\bibfield  {journal} {\bibinfo  {journal} {IEEE spectrum}\ }\textbf {\bibinfo {volume} {34}},\ \bibinfo {pages} {52} (\bibinfo {year} {1997})}\BibitemShut {NoStop}%
\bibitem [{\citenamefont {Lundstrom}(2003)}]{lundstrom2003moore}%
  \BibitemOpen
  \bibfield  {author} {\bibinfo {author} {\bibfnamefont {M.}~\bibnamefont {Lundstrom}},\ }\bibfield  {title} {\bibinfo {title} {Moore's law forever?},\ }\href@noop {} {\bibfield  {journal} {\bibinfo  {journal} {Science}\ }\textbf {\bibinfo {volume} {299}},\ \bibinfo {pages} {210} (\bibinfo {year} {2003})}\BibitemShut {NoStop}%
\bibitem [{\citenamefont {Theis}\ and\ \citenamefont {Wong}(2017)}]{theis2017end}%
  \BibitemOpen
  \bibfield  {author} {\bibinfo {author} {\bibfnamefont {T.~N.}\ \bibnamefont {Theis}}\ and\ \bibinfo {author} {\bibfnamefont {H.-S.~P.}\ \bibnamefont {Wong}},\ }\bibfield  {title} {\bibinfo {title} {The end of moore's law: A new beginning for information technology},\ }\href@noop {} {\bibfield  {journal} {\bibinfo  {journal} {Computing in science \& engineering}\ }\textbf {\bibinfo {volume} {19}},\ \bibinfo {pages} {41} (\bibinfo {year} {2017})}\BibitemShut {NoStop}%
\bibitem [{\citenamefont {Shalf}\ and\ \citenamefont {Leland}(2015)}]{shalf2015computing}%
  \BibitemOpen
  \bibfield  {author} {\bibinfo {author} {\bibfnamefont {J.~M.}\ \bibnamefont {Shalf}}\ and\ \bibinfo {author} {\bibfnamefont {R.}~\bibnamefont {Leland}},\ }\bibfield  {title} {\bibinfo {title} {Computing beyond moore's law},\ }\href@noop {} {\bibfield  {journal} {\bibinfo  {journal} {Computer}\ }\textbf {\bibinfo {volume} {48}},\ \bibinfo {pages} {14} (\bibinfo {year} {2015})}\BibitemShut {NoStop}%
\bibitem [{\citenamefont {Hirohata}\ \emph {et~al.}(2020)\citenamefont {Hirohata}, \citenamefont {Yamada}, \citenamefont {Nakatani}, \citenamefont {Prejbeanu}, \citenamefont {Di{\'e}ny}, \citenamefont {Pirro},\ and\ \citenamefont {Hillebrands}}]{hirohata2020review}%
  \BibitemOpen
  \bibfield  {author} {\bibinfo {author} {\bibfnamefont {A.}~\bibnamefont {Hirohata}}, \bibinfo {author} {\bibfnamefont {K.}~\bibnamefont {Yamada}}, \bibinfo {author} {\bibfnamefont {Y.}~\bibnamefont {Nakatani}}, \bibinfo {author} {\bibfnamefont {I.-L.}\ \bibnamefont {Prejbeanu}}, \bibinfo {author} {\bibfnamefont {B.}~\bibnamefont {Di{\'e}ny}}, \bibinfo {author} {\bibfnamefont {P.}~\bibnamefont {Pirro}},\ and\ \bibinfo {author} {\bibfnamefont {B.}~\bibnamefont {Hillebrands}},\ }\bibfield  {title} {\bibinfo {title} {Review on spintronics: Principles and device applications},\ }\href@noop {} {\bibfield  {journal} {\bibinfo  {journal} {Journal of Magnetism and Magnetic Materials}\ }\textbf {\bibinfo {volume} {509}},\ \bibinfo {pages} {166711} (\bibinfo {year} {2020})}\BibitemShut {NoStop}%
\bibitem [{\citenamefont {Naaman}\ \emph {et~al.}(2019)\citenamefont {Naaman}, \citenamefont {Paltiel},\ and\ \citenamefont {Waldeck}}]{naaman2019chiral}%
  \BibitemOpen
  \bibfield  {author} {\bibinfo {author} {\bibfnamefont {R.}~\bibnamefont {Naaman}}, \bibinfo {author} {\bibfnamefont {Y.}~\bibnamefont {Paltiel}},\ and\ \bibinfo {author} {\bibfnamefont {D.~H.}\ \bibnamefont {Waldeck}},\ }\bibfield  {title} {\bibinfo {title} {Chiral molecules and the electron spin},\ }\href@noop {} {\bibfield  {journal} {\bibinfo  {journal} {Nature Reviews Chemistry}\ }\textbf {\bibinfo {volume} {3}},\ \bibinfo {pages} {250} (\bibinfo {year} {2019})}\BibitemShut {NoStop}%
\bibitem [{\citenamefont {Quack}(1989)}]{quack1989structure}%
  \BibitemOpen
  \bibfield  {author} {\bibinfo {author} {\bibfnamefont {M.}~\bibnamefont {Quack}},\ }\bibfield  {title} {\bibinfo {title} {Structure and dynamics of chiral molecules},\ }\href@noop {} {\bibfield  {journal} {\bibinfo  {journal} {Angewandte Chemie International Edition in English}\ }\textbf {\bibinfo {volume} {28}},\ \bibinfo {pages} {571} (\bibinfo {year} {1989})}\BibitemShut {NoStop}%
\bibitem [{\citenamefont {Ray}\ \emph {et~al.}(1999)\citenamefont {Ray}, \citenamefont {Ananthavel}, \citenamefont {Waldeck},\ and\ \citenamefont {Naaman}}]{ray1999asymmetric}%
  \BibitemOpen
  \bibfield  {author} {\bibinfo {author} {\bibfnamefont {K.}~\bibnamefont {Ray}}, \bibinfo {author} {\bibfnamefont {S.}~\bibnamefont {Ananthavel}}, \bibinfo {author} {\bibfnamefont {D.}~\bibnamefont {Waldeck}},\ and\ \bibinfo {author} {\bibfnamefont {R.}~\bibnamefont {Naaman}},\ }\bibfield  {title} {\bibinfo {title} {Asymmetric scattering of polarized electrons by organized organic films of chiral molecules},\ }\href@noop {} {\bibfield  {journal} {\bibinfo  {journal} {Science}\ }\textbf {\bibinfo {volume} {283}},\ \bibinfo {pages} {814} (\bibinfo {year} {1999})}\BibitemShut {NoStop}%
\bibitem [{\citenamefont {G{\"o}hler}\ \emph {et~al.}(2011)\citenamefont {G{\"o}hler}, \citenamefont {Hamelbeck}, \citenamefont {Markus}, \citenamefont {Kettner}, \citenamefont {Hanne}, \citenamefont {Vager}, \citenamefont {Naaman},\ and\ \citenamefont {Zacharias}}]{gohler2011spin}%
  \BibitemOpen
  \bibfield  {author} {\bibinfo {author} {\bibfnamefont {B.}~\bibnamefont {G{\"o}hler}}, \bibinfo {author} {\bibfnamefont {V.}~\bibnamefont {Hamelbeck}}, \bibinfo {author} {\bibfnamefont {T.}~\bibnamefont {Markus}}, \bibinfo {author} {\bibfnamefont {M.}~\bibnamefont {Kettner}}, \bibinfo {author} {\bibfnamefont {G.}~\bibnamefont {Hanne}}, \bibinfo {author} {\bibfnamefont {Z.}~\bibnamefont {Vager}}, \bibinfo {author} {\bibfnamefont {R.}~\bibnamefont {Naaman}},\ and\ \bibinfo {author} {\bibfnamefont {H.}~\bibnamefont {Zacharias}},\ }\bibfield  {title} {\bibinfo {title} {Spin selectivity in electron transmission through self-assembled monolayers of double-stranded dna},\ }\href@noop {} {\bibfield  {journal} {\bibinfo  {journal} {Science}\ }\textbf {\bibinfo {volume} {331}},\ \bibinfo {pages} {894} (\bibinfo {year} {2011})}\BibitemShut {NoStop}%
\bibitem [{\citenamefont {Xie}\ \emph {et~al.}(2011)\citenamefont {Xie}, \citenamefont {Markus}, \citenamefont {Cohen}, \citenamefont {Vager}, \citenamefont {Gutierrez},\ and\ \citenamefont {Naaman}}]{xie2011spin}%
  \BibitemOpen
  \bibfield  {author} {\bibinfo {author} {\bibfnamefont {Z.}~\bibnamefont {Xie}}, \bibinfo {author} {\bibfnamefont {T.~Z.}\ \bibnamefont {Markus}}, \bibinfo {author} {\bibfnamefont {S.~R.}\ \bibnamefont {Cohen}}, \bibinfo {author} {\bibfnamefont {Z.}~\bibnamefont {Vager}}, \bibinfo {author} {\bibfnamefont {R.}~\bibnamefont {Gutierrez}},\ and\ \bibinfo {author} {\bibfnamefont {R.}~\bibnamefont {Naaman}},\ }\bibfield  {title} {\bibinfo {title} {Spin specific electron conduction through dna oligomers},\ }\href@noop {} {\bibfield  {journal} {\bibinfo  {journal} {Nano letters}\ }\textbf {\bibinfo {volume} {11}},\ \bibinfo {pages} {4652} (\bibinfo {year} {2011})}\BibitemShut {NoStop}%
\bibitem [{\citenamefont {Ben~Dor}\ \emph {et~al.}(2014)\citenamefont {Ben~Dor}, \citenamefont {Morali}, \citenamefont {Yochelis}, \citenamefont {Baczewski},\ and\ \citenamefont {Paltiel}}]{ben2014local}%
  \BibitemOpen
  \bibfield  {author} {\bibinfo {author} {\bibfnamefont {O.}~\bibnamefont {Ben~Dor}}, \bibinfo {author} {\bibfnamefont {N.}~\bibnamefont {Morali}}, \bibinfo {author} {\bibfnamefont {S.}~\bibnamefont {Yochelis}}, \bibinfo {author} {\bibfnamefont {L.~T.}\ \bibnamefont {Baczewski}},\ and\ \bibinfo {author} {\bibfnamefont {Y.}~\bibnamefont {Paltiel}},\ }\bibfield  {title} {\bibinfo {title} {Local light-induced magnetization using nanodots and chiral molecules},\ }\href@noop {} {\bibfield  {journal} {\bibinfo  {journal} {Nano letters}\ }\textbf {\bibinfo {volume} {14}},\ \bibinfo {pages} {6042} (\bibinfo {year} {2014})}\BibitemShut {NoStop}%
\bibitem [{\citenamefont {Kiran}\ \emph {et~al.}(2016)\citenamefont {Kiran}, \citenamefont {Mathew}, \citenamefont {Cohen}, \citenamefont {Hern{\'a}ndez~Delgado}, \citenamefont {Lacour},\ and\ \citenamefont {Naaman}}]{kiran2016helicenes}%
  \BibitemOpen
  \bibfield  {author} {\bibinfo {author} {\bibfnamefont {V.}~\bibnamefont {Kiran}}, \bibinfo {author} {\bibfnamefont {S.~P.}\ \bibnamefont {Mathew}}, \bibinfo {author} {\bibfnamefont {S.~R.}\ \bibnamefont {Cohen}}, \bibinfo {author} {\bibfnamefont {I.}~\bibnamefont {Hern{\'a}ndez~Delgado}}, \bibinfo {author} {\bibfnamefont {J.}~\bibnamefont {Lacour}},\ and\ \bibinfo {author} {\bibfnamefont {R.}~\bibnamefont {Naaman}},\ }\bibfield  {title} {\bibinfo {title} {Helicenes—a new class of organic spin filter},\ }\href@noop {} {\bibfield  {journal} {\bibinfo  {journal} {Advanced Materials}\ }\textbf {\bibinfo {volume} {28}},\ \bibinfo {pages} {1957} (\bibinfo {year} {2016})}\BibitemShut {NoStop}%
\bibitem [{\citenamefont {Eckshtain-Levi}\ \emph {et~al.}(2016)\citenamefont {Eckshtain-Levi}, \citenamefont {Capua}, \citenamefont {Refaely-Abramson}, \citenamefont {Sarkar}, \citenamefont {Gavrilov}, \citenamefont {Mathew}, \citenamefont {Paltiel}, \citenamefont {Levy}, \citenamefont {Kronik},\ and\ \citenamefont {Naaman}}]{eckshtain2016cold}%
  \BibitemOpen
  \bibfield  {author} {\bibinfo {author} {\bibfnamefont {M.}~\bibnamefont {Eckshtain-Levi}}, \bibinfo {author} {\bibfnamefont {E.}~\bibnamefont {Capua}}, \bibinfo {author} {\bibfnamefont {S.}~\bibnamefont {Refaely-Abramson}}, \bibinfo {author} {\bibfnamefont {S.}~\bibnamefont {Sarkar}}, \bibinfo {author} {\bibfnamefont {Y.}~\bibnamefont {Gavrilov}}, \bibinfo {author} {\bibfnamefont {S.~P.}\ \bibnamefont {Mathew}}, \bibinfo {author} {\bibfnamefont {Y.}~\bibnamefont {Paltiel}}, \bibinfo {author} {\bibfnamefont {Y.}~\bibnamefont {Levy}}, \bibinfo {author} {\bibfnamefont {L.}~\bibnamefont {Kronik}},\ and\ \bibinfo {author} {\bibfnamefont {R.}~\bibnamefont {Naaman}},\ }\bibfield  {title} {\bibinfo {title} {Cold denaturation induces inversion of dipole and spin transfer in chiral peptide monolayers},\ }\href@noop {} {\bibfield  {journal} {\bibinfo  {journal} {Nature communications}\ }\textbf {\bibinfo {volume} {7}},\ \bibinfo {pages} {10744} (\bibinfo {year} {2016})}\BibitemShut {NoStop}%
\bibitem [{\citenamefont {Mondal}\ \emph {et~al.}(2016)\citenamefont {Mondal}, \citenamefont {Roy}, \citenamefont {Kim}, \citenamefont {Fullerton}, \citenamefont {Cohen},\ and\ \citenamefont {Naaman}}]{mondal2016photospintronics}%
  \BibitemOpen
  \bibfield  {author} {\bibinfo {author} {\bibfnamefont {P.~C.}\ \bibnamefont {Mondal}}, \bibinfo {author} {\bibfnamefont {P.}~\bibnamefont {Roy}}, \bibinfo {author} {\bibfnamefont {D.}~\bibnamefont {Kim}}, \bibinfo {author} {\bibfnamefont {E.~E.}\ \bibnamefont {Fullerton}}, \bibinfo {author} {\bibfnamefont {H.}~\bibnamefont {Cohen}},\ and\ \bibinfo {author} {\bibfnamefont {R.}~\bibnamefont {Naaman}},\ }\bibfield  {title} {\bibinfo {title} {Photospintronics: magnetic field-controlled photoemission and light-controlled spin transport in hybrid chiral oligopeptide-nanoparticle structures},\ }\href@noop {} {\bibfield  {journal} {\bibinfo  {journal} {Nano letters}\ }\textbf {\bibinfo {volume} {16}},\ \bibinfo {pages} {2806} (\bibinfo {year} {2016})}\BibitemShut {NoStop}%
\bibitem [{\citenamefont {Alpern}\ \emph {et~al.}(2016)\citenamefont {Alpern}, \citenamefont {Katzir}, \citenamefont {Yochelis}, \citenamefont {Katz}, \citenamefont {Paltiel},\ and\ \citenamefont {Millo}}]{alpern2016unconventional}%
  \BibitemOpen
  \bibfield  {author} {\bibinfo {author} {\bibfnamefont {H.}~\bibnamefont {Alpern}}, \bibinfo {author} {\bibfnamefont {E.}~\bibnamefont {Katzir}}, \bibinfo {author} {\bibfnamefont {S.}~\bibnamefont {Yochelis}}, \bibinfo {author} {\bibfnamefont {N.}~\bibnamefont {Katz}}, \bibinfo {author} {\bibfnamefont {Y.}~\bibnamefont {Paltiel}},\ and\ \bibinfo {author} {\bibfnamefont {O.}~\bibnamefont {Millo}},\ }\bibfield  {title} {\bibinfo {title} {Unconventional superconductivity induced in nb films by adsorbed chiral molecules},\ }\href@noop {} {\bibfield  {journal} {\bibinfo  {journal} {New Journal of Physics}\ }\textbf {\bibinfo {volume} {18}},\ \bibinfo {pages} {113048} (\bibinfo {year} {2016})}\BibitemShut {NoStop}%
\bibitem [{\citenamefont {Alam}\ and\ \citenamefont {Pramanik}(2017)}]{alam2017spin}%
  \BibitemOpen
  \bibfield  {author} {\bibinfo {author} {\bibfnamefont {K.~M.}\ \bibnamefont {Alam}}\ and\ \bibinfo {author} {\bibfnamefont {S.}~\bibnamefont {Pramanik}},\ }\bibfield  {title} {\bibinfo {title} {Spin filtering with poly-t wrapped single wall carbon nanotubes},\ }\href@noop {} {\bibfield  {journal} {\bibinfo  {journal} {Nanoscale}\ }\textbf {\bibinfo {volume} {9}},\ \bibinfo {pages} {5155} (\bibinfo {year} {2017})}\BibitemShut {NoStop}%
\bibitem [{\citenamefont {Abendroth}\ \emph {et~al.}(2017)\citenamefont {Abendroth}, \citenamefont {Nakatsuka}, \citenamefont {Ye}, \citenamefont {Kim}, \citenamefont {Fullerton}, \citenamefont {Andrews},\ and\ \citenamefont {Weiss}}]{abendroth2017analyzing}%
  \BibitemOpen
  \bibfield  {author} {\bibinfo {author} {\bibfnamefont {J.~M.}\ \bibnamefont {Abendroth}}, \bibinfo {author} {\bibfnamefont {N.}~\bibnamefont {Nakatsuka}}, \bibinfo {author} {\bibfnamefont {M.}~\bibnamefont {Ye}}, \bibinfo {author} {\bibfnamefont {D.}~\bibnamefont {Kim}}, \bibinfo {author} {\bibfnamefont {E.~E.}\ \bibnamefont {Fullerton}}, \bibinfo {author} {\bibfnamefont {A.~M.}\ \bibnamefont {Andrews}},\ and\ \bibinfo {author} {\bibfnamefont {P.~S.}\ \bibnamefont {Weiss}},\ }\bibfield  {title} {\bibinfo {title} {Analyzing spin selectivity in dna-mediated charge transfer via fluorescence microscopy},\ }\href@noop {} {\bibfield  {journal} {\bibinfo  {journal} {ACS nano}\ }\textbf {\bibinfo {volume} {11}},\ \bibinfo {pages} {7516} (\bibinfo {year} {2017})}\BibitemShut {NoStop}%
\bibitem [{\citenamefont {Kumar}\ \emph {et~al.}(2017)\citenamefont {Kumar}, \citenamefont {Capua}, \citenamefont {Kesharwani}, \citenamefont {Martin}, \citenamefont {Sitbon}, \citenamefont {Waldeck},\ and\ \citenamefont {Naaman}}]{kumar2017chirality}%
  \BibitemOpen
  \bibfield  {author} {\bibinfo {author} {\bibfnamefont {A.}~\bibnamefont {Kumar}}, \bibinfo {author} {\bibfnamefont {E.}~\bibnamefont {Capua}}, \bibinfo {author} {\bibfnamefont {M.~K.}\ \bibnamefont {Kesharwani}}, \bibinfo {author} {\bibfnamefont {J.~M.}\ \bibnamefont {Martin}}, \bibinfo {author} {\bibfnamefont {E.}~\bibnamefont {Sitbon}}, \bibinfo {author} {\bibfnamefont {D.~H.}\ \bibnamefont {Waldeck}},\ and\ \bibinfo {author} {\bibfnamefont {R.}~\bibnamefont {Naaman}},\ }\bibfield  {title} {\bibinfo {title} {Chirality-induced spin polarization places symmetry constraints on biomolecular interactions},\ }\href@noop {} {\bibfield  {journal} {\bibinfo  {journal} {Proceedings of the National Academy of Sciences}\ }\textbf {\bibinfo {volume} {114}},\ \bibinfo {pages} {2474} (\bibinfo {year} {2017})}\BibitemShut {NoStop}%
\bibitem [{\citenamefont {Aragon{\`e}s}\ \emph {et~al.}(2017)\citenamefont {Aragon{\`e}s}, \citenamefont {Medina}, \citenamefont {Ferrer-Huerta}, \citenamefont {Gimeno}, \citenamefont {Teixid{\'o}}, \citenamefont {Palma}, \citenamefont {Tao}, \citenamefont {Ugalde}, \citenamefont {Giralt}, \citenamefont {D{\'\i}ez-P{\'e}rez} \emph {et~al.}}]{aragones2017measuring}%
  \BibitemOpen
  \bibfield  {author} {\bibinfo {author} {\bibfnamefont {A.~C.}\ \bibnamefont {Aragon{\`e}s}}, \bibinfo {author} {\bibfnamefont {E.}~\bibnamefont {Medina}}, \bibinfo {author} {\bibfnamefont {M.}~\bibnamefont {Ferrer-Huerta}}, \bibinfo {author} {\bibfnamefont {N.}~\bibnamefont {Gimeno}}, \bibinfo {author} {\bibfnamefont {M.}~\bibnamefont {Teixid{\'o}}}, \bibinfo {author} {\bibfnamefont {J.~L.}\ \bibnamefont {Palma}}, \bibinfo {author} {\bibfnamefont {N.}~\bibnamefont {Tao}}, \bibinfo {author} {\bibfnamefont {J.~M.}\ \bibnamefont {Ugalde}}, \bibinfo {author} {\bibfnamefont {E.}~\bibnamefont {Giralt}}, \bibinfo {author} {\bibfnamefont {I.}~\bibnamefont {D{\'\i}ez-P{\'e}rez}}, \emph {et~al.},\ }\bibfield  {title} {\bibinfo {title} {Measuring the spin-polarization power of a single chiral molecule},\ }\href@noop {} {\bibfield  {journal} {\bibinfo  {journal} {small}\ }\textbf {\bibinfo {volume} {13}},\ \bibinfo {pages} {1602519} (\bibinfo {year} {2017})}\BibitemShut {NoStop}%
\bibitem [{\citenamefont {Varade}\ \emph {et~al.}(2018)\citenamefont {Varade}, \citenamefont {Markus}, \citenamefont {Vankayala}, \citenamefont {Friedman}, \citenamefont {Sheves}, \citenamefont {Waldeck},\ and\ \citenamefont {Naaman}}]{varade2018bacteriorhodopsin}%
  \BibitemOpen
  \bibfield  {author} {\bibinfo {author} {\bibfnamefont {V.}~\bibnamefont {Varade}}, \bibinfo {author} {\bibfnamefont {T.}~\bibnamefont {Markus}}, \bibinfo {author} {\bibfnamefont {K.}~\bibnamefont {Vankayala}}, \bibinfo {author} {\bibfnamefont {N.}~\bibnamefont {Friedman}}, \bibinfo {author} {\bibfnamefont {M.}~\bibnamefont {Sheves}}, \bibinfo {author} {\bibfnamefont {D.~H.}\ \bibnamefont {Waldeck}},\ and\ \bibinfo {author} {\bibfnamefont {R.}~\bibnamefont {Naaman}},\ }\bibfield  {title} {\bibinfo {title} {Bacteriorhodopsin based non-magnetic spin filters for biomolecular spintronics},\ }\href@noop {} {\bibfield  {journal} {\bibinfo  {journal} {Physical Chemistry Chemical Physics}\ }\textbf {\bibinfo {volume} {20}},\ \bibinfo {pages} {1091} (\bibinfo {year} {2018})}\BibitemShut {NoStop}%
\bibitem [{\citenamefont {Santos}\ \emph {et~al.}(2018)\citenamefont {Santos}, \citenamefont {Rivilla}, \citenamefont {Cossio}, \citenamefont {Matxain}, \citenamefont {Grzelczak}, \citenamefont {Mazinani}, \citenamefont {Ugalde},\ and\ \citenamefont {Mujica}}]{santos2018chirality}%
  \BibitemOpen
  \bibfield  {author} {\bibinfo {author} {\bibfnamefont {J.~I.}\ \bibnamefont {Santos}}, \bibinfo {author} {\bibfnamefont {I.}~\bibnamefont {Rivilla}}, \bibinfo {author} {\bibfnamefont {F.~P.}\ \bibnamefont {Cossio}}, \bibinfo {author} {\bibfnamefont {J.~M.}\ \bibnamefont {Matxain}}, \bibinfo {author} {\bibfnamefont {M.}~\bibnamefont {Grzelczak}}, \bibinfo {author} {\bibfnamefont {S.~K.}\ \bibnamefont {Mazinani}}, \bibinfo {author} {\bibfnamefont {J.~M.}\ \bibnamefont {Ugalde}},\ and\ \bibinfo {author} {\bibfnamefont {V.}~\bibnamefont {Mujica}},\ }\bibfield  {title} {\bibinfo {title} {Chirality-induced electron spin polarization and enantiospecific response in solid-state cross-polarization nuclear magnetic resonance},\ }\href@noop {} {\bibfield  {journal} {\bibinfo  {journal} {ACS nano}\ }\textbf {\bibinfo {volume} {12}},\ \bibinfo {pages} {11426} (\bibinfo {year} {2018})}\BibitemShut {NoStop}%
\bibitem [{\citenamefont {Gazzotti}\ \emph {et~al.}(2018)\citenamefont {Gazzotti}, \citenamefont {Arnaboldi}, \citenamefont {Grecchi}, \citenamefont {Giovanardi}, \citenamefont {Cannio}, \citenamefont {Pasquali}, \citenamefont {Giacomino}, \citenamefont {Abollino},\ and\ \citenamefont {Fontanesi}}]{gazzotti2018spin}%
  \BibitemOpen
  \bibfield  {author} {\bibinfo {author} {\bibfnamefont {M.}~\bibnamefont {Gazzotti}}, \bibinfo {author} {\bibfnamefont {S.}~\bibnamefont {Arnaboldi}}, \bibinfo {author} {\bibfnamefont {S.}~\bibnamefont {Grecchi}}, \bibinfo {author} {\bibfnamefont {R.}~\bibnamefont {Giovanardi}}, \bibinfo {author} {\bibfnamefont {M.}~\bibnamefont {Cannio}}, \bibinfo {author} {\bibfnamefont {L.}~\bibnamefont {Pasquali}}, \bibinfo {author} {\bibfnamefont {A.}~\bibnamefont {Giacomino}}, \bibinfo {author} {\bibfnamefont {O.}~\bibnamefont {Abollino}},\ and\ \bibinfo {author} {\bibfnamefont {C.}~\bibnamefont {Fontanesi}},\ }\bibfield  {title} {\bibinfo {title} {Spin-dependent electrochemistry: Enantio-selectivity driven by chiral-induced spin selectivity effect},\ }\href@noop {} {\bibfield  {journal} {\bibinfo  {journal} {Electrochimica Acta}\ }\textbf {\bibinfo {volume} {286}},\ \bibinfo {pages} {271} (\bibinfo {year} {2018})}\BibitemShut {NoStop}%
\bibitem [{\citenamefont {Shapira}\ \emph {et~al.}(2018)\citenamefont {Shapira}, \citenamefont {Alpern}, \citenamefont {Yochelis}, \citenamefont {Lee}, \citenamefont {Kaun}, \citenamefont {Paltiel}, \citenamefont {Koren},\ and\ \citenamefont {Millo}}]{shapira2018unconventional}%
  \BibitemOpen
  \bibfield  {author} {\bibinfo {author} {\bibfnamefont {T.}~\bibnamefont {Shapira}}, \bibinfo {author} {\bibfnamefont {H.}~\bibnamefont {Alpern}}, \bibinfo {author} {\bibfnamefont {S.}~\bibnamefont {Yochelis}}, \bibinfo {author} {\bibfnamefont {T.-K.}\ \bibnamefont {Lee}}, \bibinfo {author} {\bibfnamefont {C.-C.}\ \bibnamefont {Kaun}}, \bibinfo {author} {\bibfnamefont {Y.}~\bibnamefont {Paltiel}}, \bibinfo {author} {\bibfnamefont {G.}~\bibnamefont {Koren}},\ and\ \bibinfo {author} {\bibfnamefont {O.}~\bibnamefont {Millo}},\ }\bibfield  {title} {\bibinfo {title} {Unconventional order parameter induced by helical chiral molecules adsorbed on a metal proximity coupled to a superconductor},\ }\href@noop {} {\bibfield  {journal} {\bibinfo  {journal} {Physical Review B}\ }\textbf {\bibinfo {volume} {98}},\ \bibinfo {pages} {214513} (\bibinfo {year} {2018})}\BibitemShut {NoStop}%
\bibitem [{\citenamefont {Ghosh}\ \emph {et~al.}(2019)\citenamefont {Ghosh}, \citenamefont {Zhang}, \citenamefont {Tassinari}, \citenamefont {Mastai}, \citenamefont {Lidor-Shalev}, \citenamefont {Naaman}, \citenamefont {Mollers}, \citenamefont {Nurenberg}, \citenamefont {Zacharias}, \citenamefont {Wei} \emph {et~al.}}]{ghosh2019controlling}%
  \BibitemOpen
  \bibfield  {author} {\bibinfo {author} {\bibfnamefont {K.}~\bibnamefont {Ghosh}}, \bibinfo {author} {\bibfnamefont {W.}~\bibnamefont {Zhang}}, \bibinfo {author} {\bibfnamefont {F.}~\bibnamefont {Tassinari}}, \bibinfo {author} {\bibfnamefont {Y.}~\bibnamefont {Mastai}}, \bibinfo {author} {\bibfnamefont {O.}~\bibnamefont {Lidor-Shalev}}, \bibinfo {author} {\bibfnamefont {R.}~\bibnamefont {Naaman}}, \bibinfo {author} {\bibfnamefont {P.}~\bibnamefont {Mollers}}, \bibinfo {author} {\bibfnamefont {D.}~\bibnamefont {Nurenberg}}, \bibinfo {author} {\bibfnamefont {H.}~\bibnamefont {Zacharias}}, \bibinfo {author} {\bibfnamefont {J.}~\bibnamefont {Wei}}, \emph {et~al.},\ }\bibfield  {title} {\bibinfo {title} {Controlling chemical selectivity in electrocatalysis with chiral cuo-coated electrodes},\ }\href@noop {} {\bibfield  {journal} {\bibinfo  {journal} {The Journal of Physical Chemistry C}\ }\textbf {\bibinfo {volume} {123}},\ \bibinfo {pages} {3024} (\bibinfo {year} {2019})}\BibitemShut {NoStop}%
\bibitem [{\citenamefont {Yeganeh}\ \emph {et~al.}(2009)\citenamefont {Yeganeh}, \citenamefont {Ratner}, \citenamefont {Medina},\ and\ \citenamefont {Mujica}}]{yeganeh2009chiral}%
  \BibitemOpen
  \bibfield  {author} {\bibinfo {author} {\bibfnamefont {S.}~\bibnamefont {Yeganeh}}, \bibinfo {author} {\bibfnamefont {M.~A.}\ \bibnamefont {Ratner}}, \bibinfo {author} {\bibfnamefont {E.}~\bibnamefont {Medina}},\ and\ \bibinfo {author} {\bibfnamefont {V.}~\bibnamefont {Mujica}},\ }\bibfield  {title} {\bibinfo {title} {Chiral electron transport: Scattering through helical potentials},\ }\href@noop {} {\bibfield  {journal} {\bibinfo  {journal} {The Journal of chemical physics}\ }\textbf {\bibinfo {volume} {131}} (\bibinfo {year} {2009})}\BibitemShut {NoStop}%
\bibitem [{\citenamefont {Guo}\ and\ \citenamefont {Sun}(2012)}]{guo2012spin}%
  \BibitemOpen
  \bibfield  {author} {\bibinfo {author} {\bibfnamefont {A.-M.}\ \bibnamefont {Guo}}\ and\ \bibinfo {author} {\bibfnamefont {Q.-f.}\ \bibnamefont {Sun}},\ }\bibfield  {title} {\bibinfo {title} {Spin-selective transport of electrons in dna double helix},\ }\href@noop {} {\bibfield  {journal} {\bibinfo  {journal} {Physical review letters}\ }\textbf {\bibinfo {volume} {108}},\ \bibinfo {pages} {218102} (\bibinfo {year} {2012})}\BibitemShut {NoStop}%
\bibitem [{\citenamefont {Gutierrez}\ \emph {et~al.}(2012)\citenamefont {Gutierrez}, \citenamefont {D{\'\i}az}, \citenamefont {Naaman},\ and\ \citenamefont {Cuniberti}}]{gutierrez2012spin}%
  \BibitemOpen
  \bibfield  {author} {\bibinfo {author} {\bibfnamefont {R.}~\bibnamefont {Gutierrez}}, \bibinfo {author} {\bibfnamefont {E.}~\bibnamefont {D{\'\i}az}}, \bibinfo {author} {\bibfnamefont {R.}~\bibnamefont {Naaman}},\ and\ \bibinfo {author} {\bibfnamefont {G.}~\bibnamefont {Cuniberti}},\ }\bibfield  {title} {\bibinfo {title} {Spin-selective transport through helical molecular systems},\ }\href@noop {} {\bibfield  {journal} {\bibinfo  {journal} {Physical Review B}\ }\textbf {\bibinfo {volume} {85}},\ \bibinfo {pages} {081404} (\bibinfo {year} {2012})}\BibitemShut {NoStop}%
\bibitem [{\citenamefont {Guo}\ and\ \citenamefont {Sun}(2014)}]{guo2014spin}%
  \BibitemOpen
  \bibfield  {author} {\bibinfo {author} {\bibfnamefont {A.-M.}\ \bibnamefont {Guo}}\ and\ \bibinfo {author} {\bibfnamefont {Q.-F.}\ \bibnamefont {Sun}},\ }\bibfield  {title} {\bibinfo {title} {Spin-dependent electron transport in protein-like single-helical molecules},\ }\href@noop {} {\bibfield  {journal} {\bibinfo  {journal} {Proceedings of the National Academy of Sciences}\ }\textbf {\bibinfo {volume} {111}},\ \bibinfo {pages} {11658} (\bibinfo {year} {2014})}\BibitemShut {NoStop}%
\bibitem [{\citenamefont {Wu}\ \emph {et~al.}(2015)\citenamefont {Wu}, \citenamefont {Zhu}, \citenamefont {Sun},\ and\ \citenamefont {Gong}}]{wu2015spin}%
  \BibitemOpen
  \bibfield  {author} {\bibinfo {author} {\bibfnamefont {H.-N.}\ \bibnamefont {Wu}}, \bibinfo {author} {\bibfnamefont {Y.-L.}\ \bibnamefont {Zhu}}, \bibinfo {author} {\bibfnamefont {X.}~\bibnamefont {Sun}},\ and\ \bibinfo {author} {\bibfnamefont {W.-J.}\ \bibnamefont {Gong}},\ }\bibfield  {title} {\bibinfo {title} {Spin polarization and spin separation realized in the double-helical molecules},\ }\href@noop {} {\bibfield  {journal} {\bibinfo  {journal} {Physica E: Low-dimensional Systems and Nanostructures}\ }\textbf {\bibinfo {volume} {74}},\ \bibinfo {pages} {156} (\bibinfo {year} {2015})}\BibitemShut {NoStop}%
\bibitem [{\citenamefont {Medina}\ \emph {et~al.}(2015)\citenamefont {Medina}, \citenamefont {Gonz{\'a}lez-Arraga}, \citenamefont {Finkelstein-Shapiro}, \citenamefont {Berche},\ and\ \citenamefont {Mujica}}]{medina2015continuum}%
  \BibitemOpen
  \bibfield  {author} {\bibinfo {author} {\bibfnamefont {E.}~\bibnamefont {Medina}}, \bibinfo {author} {\bibfnamefont {L.~A.}\ \bibnamefont {Gonz{\'a}lez-Arraga}}, \bibinfo {author} {\bibfnamefont {D.}~\bibnamefont {Finkelstein-Shapiro}}, \bibinfo {author} {\bibfnamefont {B.}~\bibnamefont {Berche}},\ and\ \bibinfo {author} {\bibfnamefont {V.}~\bibnamefont {Mujica}},\ }\bibfield  {title} {\bibinfo {title} {Continuum model for chiral induced spin selectivity in helical molecules},\ }\href@noop {} {\bibfield  {journal} {\bibinfo  {journal} {The Journal of chemical physics}\ }\textbf {\bibinfo {volume} {142}} (\bibinfo {year} {2015})}\BibitemShut {NoStop}%
\bibitem [{\citenamefont {Michaeli}\ and\ \citenamefont {Naaman}(2019)}]{michaeli2019origin}%
  \BibitemOpen
  \bibfield  {author} {\bibinfo {author} {\bibfnamefont {K.}~\bibnamefont {Michaeli}}\ and\ \bibinfo {author} {\bibfnamefont {R.}~\bibnamefont {Naaman}},\ }\bibfield  {title} {\bibinfo {title} {Origin of spin-dependent tunneling through chiral molecules},\ }\href@noop {} {\bibfield  {journal} {\bibinfo  {journal} {The Journal of Physical Chemistry C}\ }\textbf {\bibinfo {volume} {123}},\ \bibinfo {pages} {17043} (\bibinfo {year} {2019})}\BibitemShut {NoStop}%
\bibitem [{\citenamefont {Matityahu}\ \emph {et~al.}(2016)\citenamefont {Matityahu}, \citenamefont {Utsumi}, \citenamefont {Aharony}, \citenamefont {Entin-Wohlman},\ and\ \citenamefont {Balseiro}}]{matityahu2016spin}%
  \BibitemOpen
  \bibfield  {author} {\bibinfo {author} {\bibfnamefont {S.}~\bibnamefont {Matityahu}}, \bibinfo {author} {\bibfnamefont {Y.}~\bibnamefont {Utsumi}}, \bibinfo {author} {\bibfnamefont {A.}~\bibnamefont {Aharony}}, \bibinfo {author} {\bibfnamefont {O.}~\bibnamefont {Entin-Wohlman}},\ and\ \bibinfo {author} {\bibfnamefont {C.~A.}\ \bibnamefont {Balseiro}},\ }\bibfield  {title} {\bibinfo {title} {Spin-dependent transport through a chiral molecule in the presence of spin-orbit interaction and nonunitary effects},\ }\href@noop {} {\bibfield  {journal} {\bibinfo  {journal} {Physical Review B}\ }\textbf {\bibinfo {volume} {93}},\ \bibinfo {pages} {075407} (\bibinfo {year} {2016})}\BibitemShut {NoStop}%
\bibitem [{\citenamefont {Varela}\ \emph {et~al.}(2016)\citenamefont {Varela}, \citenamefont {Mujica},\ and\ \citenamefont {Medina}}]{varela2016effective}%
  \BibitemOpen
  \bibfield  {author} {\bibinfo {author} {\bibfnamefont {S.}~\bibnamefont {Varela}}, \bibinfo {author} {\bibfnamefont {V.}~\bibnamefont {Mujica}},\ and\ \bibinfo {author} {\bibfnamefont {E.}~\bibnamefont {Medina}},\ }\bibfield  {title} {\bibinfo {title} {Effective spin-orbit couplings in an analytical tight-binding model of dna: Spin filtering and chiral spin transport},\ }\href@noop {} {\bibfield  {journal} {\bibinfo  {journal} {Physical Review B}\ }\textbf {\bibinfo {volume} {93}},\ \bibinfo {pages} {155436} (\bibinfo {year} {2016})}\BibitemShut {NoStop}%
\bibitem [{\citenamefont {Pan}\ \emph {et~al.}(2016)\citenamefont {Pan}, \citenamefont {Guo},\ and\ \citenamefont {Sun}}]{pan2016spin}%
  \BibitemOpen
  \bibfield  {author} {\bibinfo {author} {\bibfnamefont {T.-R.}\ \bibnamefont {Pan}}, \bibinfo {author} {\bibfnamefont {A.-M.}\ \bibnamefont {Guo}},\ and\ \bibinfo {author} {\bibfnamefont {Q.-F.}\ \bibnamefont {Sun}},\ }\bibfield  {title} {\bibinfo {title} {Spin-polarized electron transport through helicene molecular junctions},\ }\href@noop {} {\bibfield  {journal} {\bibinfo  {journal} {Physical Review B}\ }\textbf {\bibinfo {volume} {94}},\ \bibinfo {pages} {235448} (\bibinfo {year} {2016})}\BibitemShut {NoStop}%
\bibitem [{\citenamefont {Matityahu}\ \emph {et~al.}(2017)\citenamefont {Matityahu}, \citenamefont {Aharony}, \citenamefont {Entin-Wohlman},\ and\ \citenamefont {Balseiro}}]{matityahu2017spin}%
  \BibitemOpen
  \bibfield  {author} {\bibinfo {author} {\bibfnamefont {S.}~\bibnamefont {Matityahu}}, \bibinfo {author} {\bibfnamefont {A.}~\bibnamefont {Aharony}}, \bibinfo {author} {\bibfnamefont {O.}~\bibnamefont {Entin-Wohlman}},\ and\ \bibinfo {author} {\bibfnamefont {C.~A.}\ \bibnamefont {Balseiro}},\ }\bibfield  {title} {\bibinfo {title} {Spin filtering in all-electrical three-terminal interferometers},\ }\href@noop {} {\bibfield  {journal} {\bibinfo  {journal} {Physical Review B}\ }\textbf {\bibinfo {volume} {95}},\ \bibinfo {pages} {085411} (\bibinfo {year} {2017})}\BibitemShut {NoStop}%
\bibitem [{\citenamefont {D{\'\i}az}\ \emph {et~al.}(2018{\natexlab{a}})\citenamefont {D{\'\i}az}, \citenamefont {Contreras}, \citenamefont {Hern{\'a}ndez},\ and\ \citenamefont {Dom{\'\i}nguez-Adame}}]{diaz2018effective}%
  \BibitemOpen
  \bibfield  {author} {\bibinfo {author} {\bibfnamefont {E.}~\bibnamefont {D{\'\i}az}}, \bibinfo {author} {\bibfnamefont {A.}~\bibnamefont {Contreras}}, \bibinfo {author} {\bibfnamefont {J.}~\bibnamefont {Hern{\'a}ndez}},\ and\ \bibinfo {author} {\bibfnamefont {F.}~\bibnamefont {Dom{\'\i}nguez-Adame}},\ }\bibfield  {title} {\bibinfo {title} {Effective nonlinear model for electron transport in deformable helical molecules},\ }\href@noop {} {\bibfield  {journal} {\bibinfo  {journal} {Physical Review E}\ }\textbf {\bibinfo {volume} {98}},\ \bibinfo {pages} {052221} (\bibinfo {year} {2018}{\natexlab{a}})}\BibitemShut {NoStop}%
\bibitem [{\citenamefont {D{\'\i}az}\ \emph {et~al.}(2018{\natexlab{b}})\citenamefont {D{\'\i}az}, \citenamefont {Dom{\'\i}nguez-Adame}, \citenamefont {Gutierrez}, \citenamefont {Cuniberti},\ and\ \citenamefont {Mujica}}]{diaz2018thermal}%
  \BibitemOpen
  \bibfield  {author} {\bibinfo {author} {\bibfnamefont {E.}~\bibnamefont {D{\'\i}az}}, \bibinfo {author} {\bibfnamefont {F.}~\bibnamefont {Dom{\'\i}nguez-Adame}}, \bibinfo {author} {\bibfnamefont {R.}~\bibnamefont {Gutierrez}}, \bibinfo {author} {\bibfnamefont {G.}~\bibnamefont {Cuniberti}},\ and\ \bibinfo {author} {\bibfnamefont {V.}~\bibnamefont {Mujica}},\ }\bibfield  {title} {\bibinfo {title} {Thermal decoherence and disorder effects on chiral-induced spin selectivity},\ }\href@noop {} {\bibfield  {journal} {\bibinfo  {journal} {The Journal of Physical Chemistry Letters}\ }\textbf {\bibinfo {volume} {9}},\ \bibinfo {pages} {5753} (\bibinfo {year} {2018}{\natexlab{b}})}\BibitemShut {NoStop}%
\bibitem [{\citenamefont {Maslyuk}\ \emph {et~al.}(2018)\citenamefont {Maslyuk}, \citenamefont {Gutierrez}, \citenamefont {Dianat}, \citenamefont {Mujica},\ and\ \citenamefont {Cuniberti}}]{maslyuk2018enhanced}%
  \BibitemOpen
  \bibfield  {author} {\bibinfo {author} {\bibfnamefont {V.~V.}\ \bibnamefont {Maslyuk}}, \bibinfo {author} {\bibfnamefont {R.}~\bibnamefont {Gutierrez}}, \bibinfo {author} {\bibfnamefont {A.}~\bibnamefont {Dianat}}, \bibinfo {author} {\bibfnamefont {V.}~\bibnamefont {Mujica}},\ and\ \bibinfo {author} {\bibfnamefont {G.}~\bibnamefont {Cuniberti}},\ }\bibfield  {title} {\bibinfo {title} {Enhanced magnetoresistance in chiral molecular junctions},\ }\href@noop {} {\bibfield  {journal} {\bibinfo  {journal} {The journal of physical chemistry letters}\ }\textbf {\bibinfo {volume} {9}},\ \bibinfo {pages} {5453} (\bibinfo {year} {2018})}\BibitemShut {NoStop}%
\bibitem [{\citenamefont {N{\"u}renberg}\ and\ \citenamefont {Zacharias}(2019)}]{nurenberg2019evaluation}%
  \BibitemOpen
  \bibfield  {author} {\bibinfo {author} {\bibfnamefont {D.}~\bibnamefont {N{\"u}renberg}}\ and\ \bibinfo {author} {\bibfnamefont {H.}~\bibnamefont {Zacharias}},\ }\bibfield  {title} {\bibinfo {title} {Evaluation of spin-flip scattering in chirality-induced spin selectivity using the riccati equation},\ }\href@noop {} {\bibfield  {journal} {\bibinfo  {journal} {Physical Chemistry Chemical Physics}\ }\textbf {\bibinfo {volume} {21}},\ \bibinfo {pages} {3761} (\bibinfo {year} {2019})}\BibitemShut {NoStop}%
\bibitem [{\citenamefont {Yang}\ \emph {et~al.}(2019)\citenamefont {Yang}, \citenamefont {van~der Wal},\ and\ \citenamefont {van Wees}}]{yang2019spin}%
  \BibitemOpen
  \bibfield  {author} {\bibinfo {author} {\bibfnamefont {X.}~\bibnamefont {Yang}}, \bibinfo {author} {\bibfnamefont {C.~H.}\ \bibnamefont {van~der Wal}},\ and\ \bibinfo {author} {\bibfnamefont {B.~J.}\ \bibnamefont {van Wees}},\ }\bibfield  {title} {\bibinfo {title} {Spin-dependent electron transmission model for chiral molecules in mesoscopic devices},\ }\href@noop {} {\bibfield  {journal} {\bibinfo  {journal} {Physical Review B}\ }\textbf {\bibinfo {volume} {99}},\ \bibinfo {pages} {024418} (\bibinfo {year} {2019})}\BibitemShut {NoStop}%
\bibitem [{\citenamefont {Wolf}\ \emph {et~al.}(2022)\citenamefont {Wolf}, \citenamefont {Liu}, \citenamefont {Xiao}, \citenamefont {Park},\ and\ \citenamefont {Yan}}]{wolf2022unusual}%
  \BibitemOpen
  \bibfield  {author} {\bibinfo {author} {\bibfnamefont {Y.}~\bibnamefont {Wolf}}, \bibinfo {author} {\bibfnamefont {Y.}~\bibnamefont {Liu}}, \bibinfo {author} {\bibfnamefont {J.}~\bibnamefont {Xiao}}, \bibinfo {author} {\bibfnamefont {N.}~\bibnamefont {Park}},\ and\ \bibinfo {author} {\bibfnamefont {B.}~\bibnamefont {Yan}},\ }\bibfield  {title} {\bibinfo {title} {Unusual spin polarization in the chirality-induced spin selectivity},\ }\href@noop {} {\bibfield  {journal} {\bibinfo  {journal} {ACS nano}\ }\textbf {\bibinfo {volume} {16}},\ \bibinfo {pages} {18601} (\bibinfo {year} {2022})}\BibitemShut {NoStop}%
\bibitem [{\citenamefont {Gersten}\ \emph {et~al.}(2013)\citenamefont {Gersten}, \citenamefont {Kaasbjerg},\ and\ \citenamefont {Nitzan}}]{gersten2013induced}%
  \BibitemOpen
  \bibfield  {author} {\bibinfo {author} {\bibfnamefont {J.}~\bibnamefont {Gersten}}, \bibinfo {author} {\bibfnamefont {K.}~\bibnamefont {Kaasbjerg}},\ and\ \bibinfo {author} {\bibfnamefont {A.}~\bibnamefont {Nitzan}},\ }\bibfield  {title} {\bibinfo {title} {Induced spin filtering in electron transmission through chiral molecular layers adsorbed on metals with strong spin-orbit coupling},\ }\href@noop {} {\bibfield  {journal} {\bibinfo  {journal} {The Journal of chemical physics}\ }\textbf {\bibinfo {volume} {139}} (\bibinfo {year} {2013})}\BibitemShut {NoStop}%
\bibitem [{\citenamefont {Du}\ \emph {et~al.}(2020)\citenamefont {Du}, \citenamefont {Fu},\ and\ \citenamefont {Wu}}]{du2020vibration}%
  \BibitemOpen
  \bibfield  {author} {\bibinfo {author} {\bibfnamefont {G.-F.}\ \bibnamefont {Du}}, \bibinfo {author} {\bibfnamefont {H.-H.}\ \bibnamefont {Fu}},\ and\ \bibinfo {author} {\bibfnamefont {R.}~\bibnamefont {Wu}},\ }\bibfield  {title} {\bibinfo {title} {Vibration-enhanced spin-selective transport of electrons in the dna double helix},\ }\href@noop {} {\bibfield  {journal} {\bibinfo  {journal} {Physical Review B}\ }\textbf {\bibinfo {volume} {102}},\ \bibinfo {pages} {035431} (\bibinfo {year} {2020})}\BibitemShut {NoStop}%
\bibitem [{\citenamefont {Wu}\ and\ \citenamefont {Subotnik}(2021)}]{wu2021electronic}%
  \BibitemOpen
  \bibfield  {author} {\bibinfo {author} {\bibfnamefont {Y.}~\bibnamefont {Wu}}\ and\ \bibinfo {author} {\bibfnamefont {J.~E.}\ \bibnamefont {Subotnik}},\ }\bibfield  {title} {\bibinfo {title} {Electronic spin separation induced by nuclear motion near conical intersections},\ }\href@noop {} {\bibfield  {journal} {\bibinfo  {journal} {Nature communications}\ }\textbf {\bibinfo {volume} {12}},\ \bibinfo {pages} {700} (\bibinfo {year} {2021})}\BibitemShut {NoStop}%
\bibitem [{\citenamefont {Teh}\ \emph {et~al.}(2022)\citenamefont {Teh}, \citenamefont {Dou},\ and\ \citenamefont {Subotnik}}]{teh2022spin}%
  \BibitemOpen
  \bibfield  {author} {\bibinfo {author} {\bibfnamefont {H.-H.}\ \bibnamefont {Teh}}, \bibinfo {author} {\bibfnamefont {W.}~\bibnamefont {Dou}},\ and\ \bibinfo {author} {\bibfnamefont {J.~E.}\ \bibnamefont {Subotnik}},\ }\bibfield  {title} {\bibinfo {title} {Spin polarization through a molecular junction based on nuclear berry curvature effects},\ }\href@noop {} {\bibfield  {journal} {\bibinfo  {journal} {Physical Review B}\ }\textbf {\bibinfo {volume} {106}},\ \bibinfo {pages} {184302} (\bibinfo {year} {2022})}\BibitemShut {NoStop}%
\bibitem [{\citenamefont {Nitzan}(2006)}]{nitzan2006chemical}%
  \BibitemOpen
  \bibfield  {author} {\bibinfo {author} {\bibfnamefont {A.}~\bibnamefont {Nitzan}},\ }\href@noop {} {\emph {\bibinfo {title} {Chemical dynamics in condensed phases: relaxation, transfer and reactions in condensed molecular systems}}}\ (\bibinfo  {publisher} {Oxford university press},\ \bibinfo {year} {2006})\BibitemShut {NoStop}%
\bibitem [{\citenamefont {Malhado}\ and\ \citenamefont {Hynes}(2012)}]{malhado2012photoisomerization}%
  \BibitemOpen
  \bibfield  {author} {\bibinfo {author} {\bibfnamefont {J.~P.}\ \bibnamefont {Malhado}}\ and\ \bibinfo {author} {\bibfnamefont {J.~T.}\ \bibnamefont {Hynes}},\ }\bibfield  {title} {\bibinfo {title} {Photoisomerization for a model protonated schiff base in solution: Sloped/peaked conical intersection perspective},\ }\href@noop {} {\bibfield  {journal} {\bibinfo  {journal} {The Journal of Chemical Physics}\ }\textbf {\bibinfo {volume} {137}},\ \bibinfo {pages} {22A543} (\bibinfo {year} {2012})}\BibitemShut {NoStop}%
\bibitem [{\citenamefont {Mosallanejad}\ \emph {et~al.}(2023)\citenamefont {Mosallanejad}, \citenamefont {Chen},\ and\ \citenamefont {Dou}}]{mosallanejad2023floquet}%
  \BibitemOpen
  \bibfield  {author} {\bibinfo {author} {\bibfnamefont {V.}~\bibnamefont {Mosallanejad}}, \bibinfo {author} {\bibfnamefont {J.}~\bibnamefont {Chen}},\ and\ \bibinfo {author} {\bibfnamefont {W.}~\bibnamefont {Dou}},\ }\bibfield  {title} {\bibinfo {title} {Floquet-driven frictional effects},\ }\href@noop {} {\bibfield  {journal} {\bibinfo  {journal} {Physical Review B}\ }\textbf {\bibinfo {volume} {107}},\ \bibinfo {pages} {184314} (\bibinfo {year} {2023})}\BibitemShut {NoStop}%
\bibitem [{\citenamefont {Dou}\ \emph {et~al.}(2017)\citenamefont {Dou}, \citenamefont {Miao},\ and\ \citenamefont {Subotnik}}]{dou2017born}%
  \BibitemOpen
  \bibfield  {author} {\bibinfo {author} {\bibfnamefont {W.}~\bibnamefont {Dou}}, \bibinfo {author} {\bibfnamefont {G.}~\bibnamefont {Miao}},\ and\ \bibinfo {author} {\bibfnamefont {J.~E.}\ \bibnamefont {Subotnik}},\ }\bibfield  {title} {\bibinfo {title} {Born-oppenheimer dynamics, electronic friction, and the inclusion of electron-electron interactions},\ }\href@noop {} {\bibfield  {journal} {\bibinfo  {journal} {Physical Review Letters}\ }\textbf {\bibinfo {volume} {119}},\ \bibinfo {pages} {046001} (\bibinfo {year} {2017})}\BibitemShut {NoStop}%
\bibitem [{\citenamefont {Bode}\ \emph {et~al.}(2012)\citenamefont {Bode}, \citenamefont {Kusminskiy}, \citenamefont {Egger},\ and\ \citenamefont {von Oppen}}]{bode2012current}%
  \BibitemOpen
  \bibfield  {author} {\bibinfo {author} {\bibfnamefont {N.}~\bibnamefont {Bode}}, \bibinfo {author} {\bibfnamefont {S.~V.}\ \bibnamefont {Kusminskiy}}, \bibinfo {author} {\bibfnamefont {R.}~\bibnamefont {Egger}},\ and\ \bibinfo {author} {\bibfnamefont {F.}~\bibnamefont {von Oppen}},\ }\bibfield  {title} {\bibinfo {title} {Current-induced forces in mesoscopic systems: A scattering-matrix approach},\ }\href@noop {} {\bibfield  {journal} {\bibinfo  {journal} {Beilstein Journal of Nanotechnology}\ }\textbf {\bibinfo {volume} {3}},\ \bibinfo {pages} {144} (\bibinfo {year} {2012})}\BibitemShut {NoStop}%
\bibitem [{\citenamefont {Smith}\ and\ \citenamefont {Hynes}(1993)}]{smith1993electronic}%
  \BibitemOpen
  \bibfield  {author} {\bibinfo {author} {\bibfnamefont {B.~B.}\ \bibnamefont {Smith}}\ and\ \bibinfo {author} {\bibfnamefont {J.~T.}\ \bibnamefont {Hynes}},\ }\bibfield  {title} {\bibinfo {title} {Electronic friction and electron transfer rates at metallic electrodes},\ }\href@noop {} {\bibfield  {journal} {\bibinfo  {journal} {The Journal of chemical physics}\ }\textbf {\bibinfo {volume} {99}},\ \bibinfo {pages} {6517} (\bibinfo {year} {1993})}\BibitemShut {NoStop}%
\bibitem [{\citenamefont {L{\"u}}\ \emph {et~al.}(2012)\citenamefont {L{\"u}}, \citenamefont {Brandbyge}, \citenamefont {Hedeg{\aa}rd}, \citenamefont {Todorov},\ and\ \citenamefont {Dundas}}]{lu2012current}%
  \BibitemOpen
  \bibfield  {author} {\bibinfo {author} {\bibfnamefont {J.-T.}\ \bibnamefont {L{\"u}}}, \bibinfo {author} {\bibfnamefont {M.}~\bibnamefont {Brandbyge}}, \bibinfo {author} {\bibfnamefont {P.}~\bibnamefont {Hedeg{\aa}rd}}, \bibinfo {author} {\bibfnamefont {T.~N.}\ \bibnamefont {Todorov}},\ and\ \bibinfo {author} {\bibfnamefont {D.}~\bibnamefont {Dundas}},\ }\bibfield  {title} {\bibinfo {title} {Current-induced atomic dynamics, instabilities, and raman signals: Quasiclassical langevin equation approach},\ }\href@noop {} {\bibfield  {journal} {\bibinfo  {journal} {Physical Review B}\ }\textbf {\bibinfo {volume} {85}},\ \bibinfo {pages} {245444} (\bibinfo {year} {2012})}\BibitemShut {NoStop}%
\bibitem [{\citenamefont {Chen}\ \emph {et~al.}(2018)\citenamefont {Chen}, \citenamefont {Miwa},\ and\ \citenamefont {Galperin}}]{chen2018current}%
  \BibitemOpen
  \bibfield  {author} {\bibinfo {author} {\bibfnamefont {F.}~\bibnamefont {Chen}}, \bibinfo {author} {\bibfnamefont {K.}~\bibnamefont {Miwa}},\ and\ \bibinfo {author} {\bibfnamefont {M.}~\bibnamefont {Galperin}},\ }\bibfield  {title} {\bibinfo {title} {Current-induced forces for nonadiabatic molecular dynamics},\ }\href@noop {} {\bibfield  {journal} {\bibinfo  {journal} {The Journal of Physical Chemistry A}\ }\textbf {\bibinfo {volume} {123}},\ \bibinfo {pages} {693} (\bibinfo {year} {2018})}\BibitemShut {NoStop}%
\bibitem [{\citenamefont {Dou}\ \emph {et~al.}(2015)\citenamefont {Dou}, \citenamefont {Nitzan},\ and\ \citenamefont {Subotnik}}]{dou2015frictional}%
  \BibitemOpen
  \bibfield  {author} {\bibinfo {author} {\bibfnamefont {W.}~\bibnamefont {Dou}}, \bibinfo {author} {\bibfnamefont {A.}~\bibnamefont {Nitzan}},\ and\ \bibinfo {author} {\bibfnamefont {J.~E.}\ \bibnamefont {Subotnik}},\ }\bibfield  {title} {\bibinfo {title} {Frictional effects near a metal surface},\ }\href@noop {} {\bibfield  {journal} {\bibinfo  {journal} {The Journal of chemical physics}\ }\textbf {\bibinfo {volume} {143}},\ \bibinfo {pages} {054103} (\bibinfo {year} {2015})}\BibitemShut {NoStop}%
\bibitem [{\citenamefont {Teh}\ \emph {et~al.}(2021)\citenamefont {Teh}, \citenamefont {Dou},\ and\ \citenamefont {Subotnik}}]{teh2021antisymmetric}%
  \BibitemOpen
  \bibfield  {author} {\bibinfo {author} {\bibfnamefont {H.-H.}\ \bibnamefont {Teh}}, \bibinfo {author} {\bibfnamefont {W.}~\bibnamefont {Dou}},\ and\ \bibinfo {author} {\bibfnamefont {J.~E.}\ \bibnamefont {Subotnik}},\ }\bibfield  {title} {\bibinfo {title} {Antisymmetric berry frictional force at equilibrium in the presence of spin-orbit coupling},\ }\href@noop {} {\bibfield  {journal} {\bibinfo  {journal} {Physical Review B}\ }\textbf {\bibinfo {volume} {104}},\ \bibinfo {pages} {L201409} (\bibinfo {year} {2021})}\BibitemShut {NoStop}%
\bibitem [{\citenamefont {Haug}\ \emph {et~al.}(2008)\citenamefont {Haug}, \citenamefont {Jauho} \emph {et~al.}}]{haug2008quantum}%
  \BibitemOpen
  \bibfield  {author} {\bibinfo {author} {\bibfnamefont {H.}~\bibnamefont {Haug}}, \bibinfo {author} {\bibfnamefont {A.-P.}\ \bibnamefont {Jauho}}, \emph {et~al.},\ }\href@noop {} {\emph {\bibinfo {title} {Quantum kinetics in transport and optics of semiconductors}}},\ Vol.~\bibinfo {volume} {2}\ (\bibinfo  {publisher} {Springer},\ \bibinfo {year} {2008})\BibitemShut {NoStop}%
\bibitem [{\citenamefont {Meir}\ and\ \citenamefont {Wingreen}(1992)}]{meir1992landauer}%
  \BibitemOpen
  \bibfield  {author} {\bibinfo {author} {\bibfnamefont {Y.}~\bibnamefont {Meir}}\ and\ \bibinfo {author} {\bibfnamefont {N.~S.}\ \bibnamefont {Wingreen}},\ }\bibfield  {title} {\bibinfo {title} {Landauer formula for the current through an interacting electron region},\ }\href@noop {} {\bibfield  {journal} {\bibinfo  {journal} {Physical review letters}\ }\textbf {\bibinfo {volume} {68}},\ \bibinfo {pages} {2512} (\bibinfo {year} {1992})}\BibitemShut {NoStop}%
\bibitem [{\citenamefont {Gutierrez}\ \emph {et~al.}(2013)\citenamefont {Gutierrez}, \citenamefont {D{\'\i}az}, \citenamefont {Gaul}, \citenamefont {Brumme}, \citenamefont {Dom{\'\i}nguez-Adame},\ and\ \citenamefont {Cuniberti}}]{gutierrez2013modeling}%
  \BibitemOpen
  \bibfield  {author} {\bibinfo {author} {\bibfnamefont {R.}~\bibnamefont {Gutierrez}}, \bibinfo {author} {\bibfnamefont {E.}~\bibnamefont {D{\'\i}az}}, \bibinfo {author} {\bibfnamefont {C.}~\bibnamefont {Gaul}}, \bibinfo {author} {\bibfnamefont {T.}~\bibnamefont {Brumme}}, \bibinfo {author} {\bibfnamefont {F.}~\bibnamefont {Dom{\'\i}nguez-Adame}},\ and\ \bibinfo {author} {\bibfnamefont {G.}~\bibnamefont {Cuniberti}},\ }\bibfield  {title} {\bibinfo {title} {Modeling spin transport in helical fields: derivation of an effective low-dimensional hamiltonian},\ }\href@noop {} {\bibfield  {journal} {\bibinfo  {journal} {The Journal of Physical Chemistry C}\ }\textbf {\bibinfo {volume} {117}},\ \bibinfo {pages} {22276} (\bibinfo {year} {2013})}\BibitemShut {NoStop}%
\bibitem [{\citenamefont {Chen}\ \emph {et~al.}(2023)\citenamefont {Chen}, \citenamefont {Liu},\ and\ \citenamefont {Dou}}]{chen2023floquet}%
  \BibitemOpen
  \bibfield  {author} {\bibinfo {author} {\bibfnamefont {J.}~\bibnamefont {Chen}}, \bibinfo {author} {\bibfnamefont {W.}~\bibnamefont {Liu}},\ and\ \bibinfo {author} {\bibfnamefont {W.}~\bibnamefont {Dou}},\ }\bibfield  {title} {\bibinfo {title} {Floquet nonadiabatic nuclear dynamics with photoinduced lorenz-like force in quantum transport},\ }\href@noop {} {\bibfield  {journal} {\bibinfo  {journal} {arXiv preprint arXiv:2308.12660}\ } (\bibinfo {year} {2023})}\BibitemShut {NoStop}%
\bibitem [{\citenamefont {Evers}\ \emph {et~al.}(2022)\citenamefont {Evers}, \citenamefont {Aharony}, \citenamefont {Bar-Gill}, \citenamefont {Entin-Wohlman}, \citenamefont {Hedeg{\aa}rd}, \citenamefont {Hod}, \citenamefont {Jelinek}, \citenamefont {Kamieniarz}, \citenamefont {Lemeshko}, \citenamefont {Michaeli} \emph {et~al.}}]{evers2022theory}%
  \BibitemOpen
  \bibfield  {author} {\bibinfo {author} {\bibfnamefont {F.}~\bibnamefont {Evers}}, \bibinfo {author} {\bibfnamefont {A.}~\bibnamefont {Aharony}}, \bibinfo {author} {\bibfnamefont {N.}~\bibnamefont {Bar-Gill}}, \bibinfo {author} {\bibfnamefont {O.}~\bibnamefont {Entin-Wohlman}}, \bibinfo {author} {\bibfnamefont {P.}~\bibnamefont {Hedeg{\aa}rd}}, \bibinfo {author} {\bibfnamefont {O.}~\bibnamefont {Hod}}, \bibinfo {author} {\bibfnamefont {P.}~\bibnamefont {Jelinek}}, \bibinfo {author} {\bibfnamefont {G.}~\bibnamefont {Kamieniarz}}, \bibinfo {author} {\bibfnamefont {M.}~\bibnamefont {Lemeshko}}, \bibinfo {author} {\bibfnamefont {K.}~\bibnamefont {Michaeli}}, \emph {et~al.},\ }\bibfield  {title} {\bibinfo {title} {Theory of chirality induced spin selectivity: Progress and challenges},\ }\href@noop {} {\bibfield  {journal} {\bibinfo  {journal} {Advanced Materials}\ }\textbf {\bibinfo {volume} {34}},\ \bibinfo {pages} {2106629} (\bibinfo {year} {2022})}\BibitemShut {NoStop}%
\end{thebibliography}%

\end{document}